\documentclass[aps,prb,reprint,superscriptaddress,longbibliography]{revtex4-2}
\usepackage{graphicx}
\usepackage{textcomp}
\usepackage{gensymb}
\usepackage{amsmath, amssymb}
\usepackage[colorlinks = true, citecolor = magenta]{hyperref}
\usepackage{braket}
\usepackage{natbib}
\usepackage[utf8]{inputenc}
\usepackage{float}
\usepackage{graphicx}
\usepackage{overpic} 

\begin{document}

\title{Homodyne measurement of a non-Hermitian qubit undergoing spontaneous emission}

\author{Roson~Nongthombam}
\email{n.roson@iitg.ac.in}
\affiliation{Department of Physics, Indian Institute of Technology Guwahati, Guwahati-781039, (India)}

\author{Amarendra~K.~Sarma}
\email{aksarma@iitg.ac.in}
\affiliation{Department of Physics, Indian Institute of Technology Guwahati, Guwahati-781039, (India)}

\date{\today}

\begin{abstract}
Implementation of a two-level non-Hermitian qubit via postselection of a three-level 
system has been demonstrated. 
The post-selection procedure, which discards quantum jump to the ground-state manifold 
$\lvert g \rangle$ while retaining excitations in the $\lvert e \rangle$ and 
$\lvert f \rangle$ manifolds, effectively generates a non-Hermitian qubit exhibiting 
$\mathcal{PT}$ symmetry. 
In this work, we perform continuous homodyne measurement of this non-Hermitian qubit 
and analyze the interplay between decay introduced by post-selection and measurement backaction. 
We compare the ensemble-averaged dynamics obtained from measurement trajectories 
with the Liouvillian average, highlighting the deviations arising from measurement backaction and postselection.
We formulate the no-jump stochastic differential equation describing the postselected non-Hermitian qubit and show that its ensemble-averaged dynamics agree with those of the jump-updated postselected evolution at drive strengths far from the Liouvillian exceptional point (EP).
The degree of deviation near the EP depends sensitively on the nature of the drive.
This discrepancy is attributed to the interplay between measurement backaction and the non-Hermitian decay introduced by post-selection.
Furthermore, we determine the optimal path of the non-Hermitian qubit by extremizing the action within the path-integral formulation of the quantum trajectory framework.
Our results provide insights into how measurement backaction and non-Hermitian dynamics 
together shape the transient behavior of open quantum systems and enable controlled 
manipulation of qubits near exceptional points. 

\end{abstract}

\maketitle
\section{Introduction}

Unlike a Hermitian quantum system, where the evolution is coherent, unitary, and trace-preserving, the evolution of a non-Hermitian system is still coherent but non-unitary and non–trace-preserving.
Such systems inherently include loss and gain channels, and when driven, they can exhibit parity–time (\(\mathcal{PT}\)) symmetry breaking transitions, where the energy spectrum changes from being purely real to imaginary \cite{Bender_2007, Cbender_1999, PhysRevA_91_052113}. 
The transition point is known as the exceptional point (EP) of the non-Hermitian system. 
A non-Hermitian system with only loss or only gain channels can also exhibit an EP.
At the EP, the eigenvalues coalesce and the corresponding eigenvectors become parallel (i.e., linearly dependent) \cite{heiss2012, Miri_2019,Ozdemir_2019}. 
Non-Hermitian systems have been realized in various classical platforms \cite{Peng2014, feng2017, hodaei2014, PhysRevLett.115.040402, xiao2017, peng}, and transitions in the vicinity of EPs have been exploited in a range of applications \cite{hodaei2017, xu2016, zhang2017, PhysRevB.100.134505, shi2016, chen2017, lau2018}.

In a generic two-level system, non Hermiticity and the EP can be introduced either by making one of the levels lossy or amplifying (gain), or by assigning loss to one level and gain to the other, while keeping the two-levels coherently coupled. 
In the latter case, the system exhibits \(\mathcal{PT}\) symmetry. 
In both scenarios, when the coupling strength between the two-levels becomes comparable to the loss or gain, the system reaches an EP.
For the \(\mathcal{PT}\) symmetric case, where one level experiences gain and the other loss, the system supports coherent oscillations of population between the two-levels in the unbroken \(\mathcal{PT}\) symmetric phase. 
In contrast, in the broken \(\mathcal{PT}\) symmetric phase, the population decays exponentially without oscillations \cite{naghiloo2019}.

Here we consider a two-level non-Hermitian system constructed via post-selection on a three-level system, following the approach studied in a superconducting qubit platform \cite{naghiloo2019,  BliasRevModPhys.93.025005, KochPhysRevA.76.042319}. 
The post-selection is carried out by discarding all measurement outcomes in which the system goes to the ground state manifold $\lvert g \rangle$, and retaining only those in which the population remains in the first excited state $\lvert e \rangle$ or the second excited state $\lvert f \rangle$ manifold \cite{naghiloo2019}. 
This effectively reduces the three-level system to a non Hermitian two-level system, or a non Hermitian qubit.  
This system exhibits \(\mathcal{PT}\) symmetry, where the dynamics are governed by the difference between the eigenvalues. 
If spontaneous emission or quantum jumps of population between the levels are included, the degeneracy of eigenvalues at the EP is lifted, and the system exhibits a second order exceptional point \cite{Ming2019, Chen2021}. 
In the unbroken \(\mathcal{PT}\) symmetric phase, the dynamics show decaying oscillations, whereas in the broken \(\mathcal{PT}\) symmetric phase, the system becomes overdamped. 
The inclusion of quantum jumps implies that the presence of an EP can only be observed in the transient dynamics of the system \cite{Ming2019}. 
This is because the eigenstates corresponding to the EP are unphysical in the steady state: the system always decays into the eigenstate associated with the smallest decay rate , which does not correspond to the EP \cite{Ming2019, Chen2021, PhysRevA.101.062112}.
In this work, we study the homodyne measurement of the non-Hermitian qubit undergoing spontaneous emission. 

The unraveling of quantum trajectories of a quantum system under continuous measurement using diffusive and photo-counting methods has been widely studied \cite{hmwiseman1996, steck2007quantum, PhysRevX.6.011002, murch2013}. In particular, continuous homodyne measurement of a simple two-level system undergoing spontaneous emission has been extensively investigated \cite{lewalle2020, jordan2016, PhysRevA.96.053807, PhysRevX.6.011002, naghiloo2016, PhysRevA.96.022104, ficheux2018}. Similarly, the photo-counting method has also been studied \cite{lewalle2020, PhysRevResearch.6.L032057}.
The theoretical foundation for describing quantum trajectories lies in quantum measurement theory, which is a well-established field \cite{hmwiseman1996, steck2007quantum, wiseman2009, PhysRevLett.70.548, weber2014, jordan2013, mlmer1993, PhysRevA.89.023827, jacobs2006, PhysRevA.94.042326, murch2013, minev2019, PhysRevA.77.012112, PhysRevLett.52.1657, PhysRevA.34.1642, PhysRevA.96.022104,etde_21197676,PhysRevLett.123.163601,Breuer_lecture_notes,RivasHuelga2012,Carmichael1993,Lami_2024}.
Here, we use both schemes to construct a hybrid quantum detection protocol that unravels the quantum trajectories of a fluorescent three-level system. Using this hybrid measurement scheme, we study the quantum diffusive measurement of a non-Hermitian qubit obtained via postselection and investigate the effect of the nonlinearity induced by postselection. In this approach, photons emitted from the qubit are continuously collected and monitored, and the qubit state is inferred from the stochastic trajectories generated by the measurement record.
By performing the measurement, we investigate the interplay between non-Hermitian decay, which is referred to as the decay introduced by post-selection or decay to the ground state without a quantum jump, and the measurement backaction arising from homodyne detection.
We then derive the no-jump stochastic differential equation describing the postselected non-Hermitian qubit and compare its dynamics with the Liouvillian evolution.
The ensemble-averaged dynamics from this no-jump equation agree with the Liouvillian at drive strengths far from the exceptional point (EP).
Near the EP, the deviation is dependent on the axis of the drive.
This dependence is evident in whether the system is driven about the $x$-axis or the $y$-axis of the qubit’s Bloch vector. 
We further show the optimal path of the measurement process by incorporating the path integral formalism in the quantum trajectory framework \cite{PhysRevA.88.042110, PhysRevA.92.032125, PhysRevA.95.042126, PRXQuantum.3.010327, PhysRevA.98.012141}.
\vspace{-2mm}

The paper is organized as follows. In Section II, we discuss different measurement setups that unravel quantum trajectories. We define the Kraus operators associated with these setups and analyze the resulting trajectories. Section III focuses on the post-selection of trajectories and the definition of the non-Hermitian qubit. In Section IV, we compare the dynamics of the qubit obtained through two approaches: one based on the Liouvillian restricted to the qubit manifold, and the other derived from averaging the postselected trajectories. In Section V, we derive the stochastic master equation for the non-Hermitian qubit under homodyne detection and study the combined effects of measurement backaction and post-selection. Section VI presents the optimal paths of the qubit using a path-integral formalism. Finally, in Section VII, we provide a brief summary of the work.

\section{Hybrid Measurement Setup of the Three-Level System}
\label{Hybrid Measurement Setup of the Three-Level System}

\begin{figure}[ht]
    \centering
    \begin{overpic}[width=0.35\textwidth]{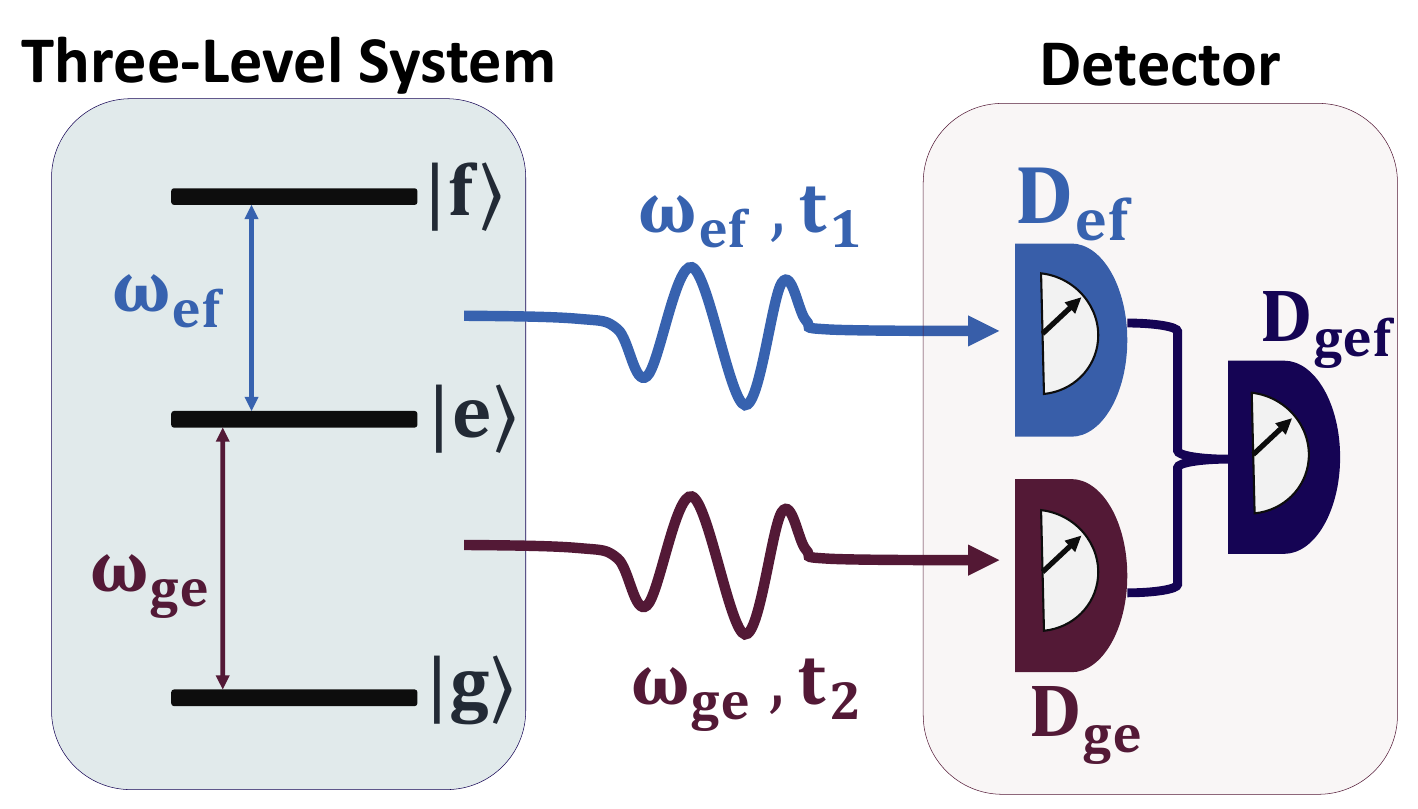}
        \put(-10,60){(a)}
    \end{overpic}

    \hspace{3mm}

    \begin{overpic}[width=0.43\textwidth]{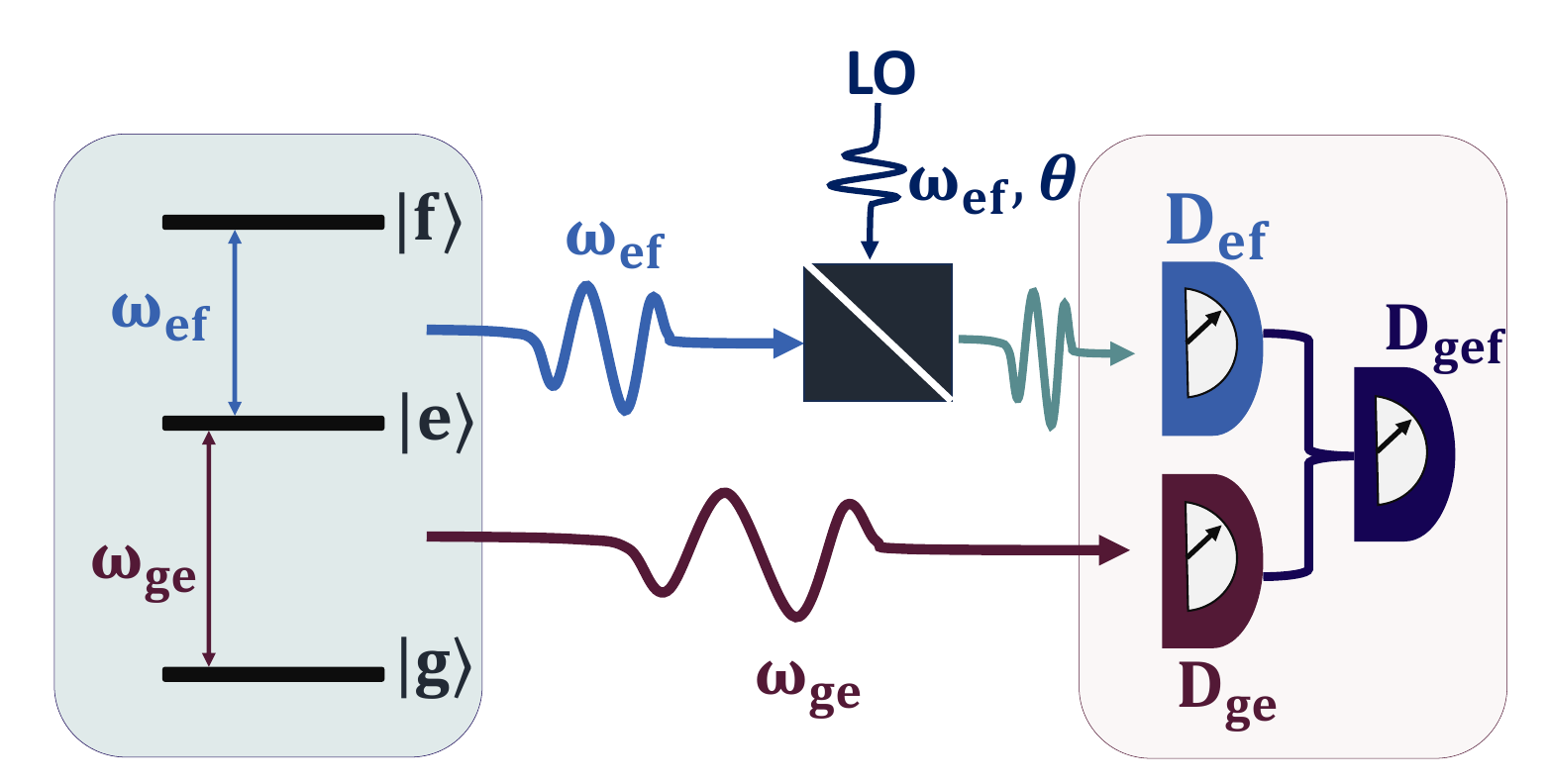}
        \put(0,47){(b)}
    \end{overpic}
    \caption{Measurement schemes of the three-level system illustrating different unraveling models of its trajectories. The system emits photons that are directed to the detector via the environment or a cavity. The detector has three components. One part ($D_{ef}$) detects photons released from the $\lvert f \rangle \to \lvert e \rangle$ transition with frequency $\omega_{ef}$, and another ($D_{eg}$) detects photons from the $\lvert e \rangle \to \lvert g \rangle$ transition with frequency $\omega_{eg}$. The detected signals from these two components are sent to $D_{gef}$, where the final state update happens. The schematic shows two types of measurement. (a) A pure photon-counting detection: at some time $t_1$, when a jump from $\lvert f \rangle$ to $\lvert e \rangle$ occurs, $D_{ef}$ clicks and at some other time $t_2$, when a jump from $\lvert e \rangle$ to $\lvert g \rangle$ occurs, $D_{eg}$ clicks. In the former case, the detector $D_{gef}$ updates the state according to the Kraus operator $K_{1e}$, and in the latter according to $K_{1g}$. When no jump occurs, the state is always updated according to $K_0$. (b) A hybrid measurement scheme: the jump from $\lvert f \rangle$ to $\lvert e \rangle$ is continuously and weakly monitored through a homodyne measurement setup, while the jump from $\lvert e \rangle$ to $\lvert g \rangle$ is monitored by photon counting measurement. The weak and continuous homodyne measurement is realized by inserting a beam splitter in front of the detector, and a local oscillator of frequency $\omega_{ef}$ and phase $\theta$ (with respect to the $\lvert f \rangle \to \lvert e \rangle$ photon) is sent to the beam splitter. We assume that, because of the frequency mismatch, the $\lvert e \rangle \to \lvert g \rangle$ photon is sent to the photon-counting channel. As long as there is no jump from $\lvert e \rangle$ to $\lvert g \rangle$, the detector $D_{gef}$ updates the state as per $K_h$, and upon detection it updates via $K_J$.}
    \label{fig:System}
\end{figure}

Let us consider the measurement of a generic three-level system which interacts with an environment. 
Assuming no interaction between the system and the environment initially, the state of the system plus environment is given as
\begin{equation}
    \ket{\Psi(0)} = \left( c_g \ket{g} + c_e \ket{e} + c_f \ket{f} \right) \otimes \ket{E_0},
\end{equation}
where $\ket{f}$, $\ket{e}$, and $\ket{g}$ are the second excited state, first excited state, and ground state, respectively. $\ket{E_0}$ is the initial state of the environment. 
The quantities $|c_g|^2$, $|c_e|^2$, and $|c_f|^2$ represent the probabilities of finding the system in the states $\ket{g}$, $\ket{e}$, and $\ket{f}$, respectively.
The system interacts with the environment and undergoes spontaneous emission from $\ket{f}$ to $\ket{e}$ or from $\ket{e}$ to $\ket{g}$, all of which have different transition frequencies. 
There is no direct transition from $\ket{f}$ to $\ket{g}$.
The combined system-environment state evolves as \(\ket{\Psi(t)} = U_{t,0}\ket{\Psi(0)}\), where \(U_{t,0}\) is the unitary operator that evolves the joint system--environment state from the initial time \(t = 0\) to time \(t\), and \(\ket{\Psi(0)}\) is given by Eq.~(1). Suppose that the environment is projected onto one of its eigenstates \(\ket{E_i}\). Then, the state of the system is correspondingly projected to \(\ket{\Psi(t)}_s = K_i \ket{\Psi(0)}_s\), where \(K_i = \bra{E_i} U_{t,0} \ket{E_0}\) is the Kraus operator acting on the system, associated with the measurement outcome corresponding to the environment state \(\ket{E_i}\). Instead of specifying the exact form of \(U_{t,0}\) and determining the state at an arbitrary time \(t\), we phenomenologically describe the evolved state as
\begin{align}
    \ket{\Psi(t + dt)} =\; & c_e(t) \sqrt{p_g} \ket{g} \ket{E_{1g}} + c_f(t) \sqrt{p_e} \ket{e} \ket{E_{1e}} \nonumber \\
    & + \left( c_g(t) \ket{g} + c_e(t) \sqrt{1-p_g} \ket{e} \right. \nonumber \\
    & \left. + c_f(t) \sqrt{1-p_e} \ket{f} \right) \ket{E_0},
\end{align}

where $\ket{E_0}$ is the state of the environment when no photon is emitted from the system, whereas $\ket{E_{1g}}$ and $\ket{E_{1e}}$ are the environment states when the system emits a photon from $\ket{e}$ to $\ket{g}$ with probability $|c_e|^2p_g $ and from $\ket{f}$ to $\ket{e}$ with probability $|c_f|^2p_e $, respectively. Here, $p_g = \gamma_g dt$ and $p_e = \gamma_e dt$ where $ \gamma_e = 1/T_e$ and $\gamma_g = 1/T_g$ are the characteristic rates at which the system undergoes spontaneous emission.

The emitted photon from the system to the environment during the spontaneous emission is directed to a detector as shown in Fig.\ref{fig:System}. 
As discussed in the figure, we assume that the detector is able to distinguish between the $\ket{f} \to \ket{e}$ transition photon and the $\ket{e} \to \ket{g}$ photon. 
We also assume that $dt \ll T_e, T_g$ (i.e., $\gamma_e dt \ll 1$ and $\gamma_g dt \ll 1$), so that the detector is Markovian.
When no photon is emitted from the system during the interval \(t\) to \(t+dt\), the detector measures the environment to be in the state \( |E_0\rangle \), thereby projecting the system state to \( |\Psi(t+dt)\rangle_s = \langle E_0 | \Psi(t+dt) \rangle = K_0 |\Psi(t)\rangle_s \), where $K_0$ is the corresponding Kraus operator and $\ket{\Psi(t)}_s = c_g \ket{g} + c_e \ket{e} + c_f \ket{f}$ is the state of the three-level system. 
In contrast, if a photon is emitted due to the transition $\ket{f} \to \ket{e}$, the detector measures the environment to be in the state $\ket{E_{1e}}$, giving $\braket{E_{1e} | \Psi(t+dt)} = K_{1e} \ket{\Psi(t)}_s$ as the state of the system, with $K_{1e}$ as the associated Kraus operator. Similarly, when photon emission arises from the transition \( |e\rangle \to |g\rangle \), the measurement outcome corresponding to the environment state \( |E_{1g}\rangle \) projects the system to \( \langle E_{1g} | \Psi(t+dt) \rangle = K_{1g} |\Psi(t)\rangle_s \). where $K_{1g}$ represents the corresponding Kraus operator for that process.
The three Kraus operators can be obtained from the Kraus matrix
\begin{equation}
    M = \begin{bmatrix}
    \sqrt{1-p_e} & 0 & 0 \\
    \sqrt{p_e} \hat{a}^\dagger_e & \sqrt{1-p_g} & 0 \\
    0 & \sqrt{p_g} \hat{a}^\dagger_g & 1
    \end{bmatrix},
\end{equation}
Here, the Kraus operators are defined as $K_0 = \braket{E_0 | M | E_0}$, $K_{1g} = \braket{E_{1g} | M | E_0}$, and $K_{1e} = \braket{E_{1e} | M | E_0}$, where the creation operators act on the environment state as $\hat{a}^\dagger_e \ket{E_0} = \ket{E_{1e}}$ and $\hat{a}^\dagger_g \ket{E_0} = \ket{E_{1g}}$, and the environment states are assumed to be orthonormal. Note that this measurement scheme illustrates one of the unraveling models for the trajectories in a three-level measurement.

We now consider another trajectory unraveling model using a hybrid detection scheme, which continuously monitors the $\ket{f} \to \ket{e}$ transition diffusively while simultaneously photon-counting the $\ket{e} \to \ket{g}$ jump, as shown in Fig.~\ref{fig:System}(b).
Since the system has different transition frequencies, we assume that the hybrid detector can distinguish between the emissions and is able to use different measurement setups for the different transitions as discussed in the figure. 
The hybrid detector continuously monitors the $\ket{f} \to \ket{e}$ quantum jump via weak homodyne detection, a phase-sensitive technique where the signal is mixed with a strong local oscillator to measure a specific field quadrature in time, as shown in Fig.~\ref{fig:System}(b).
When a jump from $\ket{e}$ to $\ket{g}$ occurs, the population in $\ket{e}$ and $\ket{f}$ vanishes, the system collapses into $\ket{g}$. The state update of this detection process can be described using the Kraus operators of the detector $D_\text{gef}$.

The Kraus operators \(K_h\) and \(K_J\), which operate on the system when the measurement is homodyne and photon-counting, respectively, are obtained from the relations
\(
K_h = \langle X | M | E_0 \rangle, 
\) and \(
K_J = \langle E_{1g} | M | E_0 \rangle,
\)
where \(X\) is the measured quadrature in the homodyne measurement. In terms of the measurement outcome \(r\) of the homodyne detection, this quadrature can be written as
\(
X = r \sqrt{\frac{dt}{2}}
\) \cite{lewallethesis}.
Using the projections
\begin{align}
    \langle X | E_0 \rangle &= \left( \frac{1}{\pi} \right)^{1/4} e^{-X^2/2}, \\
    \langle X | E_{1e} \rangle &= \left( \frac{1}{\pi} \right)^{1/4} e^{-X^2/2} \sqrt{2}\, X, \\
    \langle X | E_{1g} \rangle &= 0,
\end{align}
the Kraus operators can be written explicitly as follows.
When no jump is registered, the operator is
\begin{equation}
    K_h = \sqrt{N} e^{-r^2 dt / 4}
    \begin{bmatrix}
    \sqrt{1 - \gamma_e dt} & 0 & 0 \\
    r \, dt \sqrt{\gamma_e} e^{-i \theta} & \sqrt{1 - \gamma_g dt} & 0 \\
    0 & 0 & 1
    \end{bmatrix},
\end{equation}
where \(\theta\) is the phase difference between the \(\ket{f} \to \ket{e}\) transition and the local oscillator.
When a jump is registered, the operator is
\begin{equation}
    K_J = \sqrt{N} e^{-r^2 dt / 4}
    \begin{bmatrix}
    0 & 0 & 0 \\
    0 & 0 & 0 \\
    0 & \sqrt{\gamma_g dt} & 0
    \end{bmatrix}.
\end{equation}
The measurement Kraus operators \(K_h\) and \(K_J\) satisfy the POVM condition,
\(
\int dr ( K_h^\dagger K_h + K_J^\dagger K_J ) = \mathbb{I},
\)
which yields the normalization factor
\(
N = \sqrt{\frac{dt}{2\pi}}.
\)

To numerically simulate the state-update trajectories of the measurement, we compute the relative probabilities of whether the \(\ket{e} \to \ket{g}\) jump is registered by the detector.
The jump is registered with probability
\(\
P_J = \mathrm{tr} \left( K_J \, \rho(t) \, K_J^\dagger \right),
\)
which, upon expansion, gives $P_J = w_J\times g_J$ where
\begin{equation}
w_J = \rho_{ee}\gamma_g dt\, e^{-\frac{\gamma_g}{2}dt}
\quad \text{and} \quad
g_J = \sqrt{\frac{dt}{2\pi}}\, e^{-\frac{dt}{2}r^2}.
\label{eq:wJgJ}
\end{equation}

No jump occurs with probability
$
P_h = \mathrm{tr} \left( K_h \, \rho(t) \, K_h^\dagger \right),
$
which, upon expansion, gives $P_h = w_h\times g_h$ where
\begin{equation}
    w_h =\;  exp\left[ \left( \frac{(e^{i\theta}\rho_{ef} +e^{-i\theta}\rho_{ef})^2}{2(\rho_{ff}+\rho_{ee} + \rho_{gg})} - \gamma_e \rho_{ff} - \gamma_g \rho_{ee} \right)dt\right],
    \label{eq:wh}
\end{equation}
and 
\begin{equation}
    g_h =\; \sqrt{\frac{dt}{2\pi}} \;\,\, exp\left[ -\frac{dt}{2}\left( r - \frac{\sqrt{\gamma_e}(e^{i\theta}\rho_{ef} +e^{-i\theta}\rho_{fe})}{(\rho_{ff}+\rho_{ee} + \rho_{gg})}  \right)^2\right].
    \label{eq:gh}
\end{equation}

Both probabilities are separated into weight factors \(w_h\) and \(w_J\), multiplied by terms \(g_h\) and \(g_J\), which are Gaussian in \(r\).
In the no-jump case, the homodyne record \(r\) is inferred from a Gaussian distribution \(g_h\) with mean
\( \sqrt{\gamma_e} \left( (\rho_{ef} e^{i\theta} + \rho_{fe} e^{-i\theta})/(\rho_{ff} + \rho_{ee} + \rho_{gg}) \right). \)
In the jump case, the homodyne record \(r\) is inferred from a Gaussian distribution \(g_J\) with mean \(0\).
The two weight factors are normalized as
\(
W_h = w_h/(w_h + w_J), \)
and 
\(
W_J = w_J/(w_h + w_J), \)
such that \(W_h + W_J = 1\).
A binomial distribution is then sampled using these normalized weights: outcome \(0\) corresponds to no jump with probability \(W_h\), and outcome \(1\) corresponds to a jump with probability \(W_J\).
Whenever the outcome is \(0\), the system state is updated using \(\rho_h(t+dt)\), and when the outcome is \(1\), it is updated using \(\rho_J(t+dt)\).
The state update of the system, depending on whether a jump is registered or not, is given by
\begin{equation}
    \rho(t + dt)|_{h,J} = \frac{U K_{h,J} \, \rho(t) \, K_{h,J}^\dagger U^\dagger}{\mathrm{tr} \left( U K_{h,J} \, \rho(t) \, K_{h,J}^\dagger U^\dagger \right)},
    \label{eq:state_uodate}
\end{equation}
where $U$ is the unitary evolution operator of the system and the subscript $h$ or $J$ corresponds to the homodyne or jump measurement, respectively. 

\begin{figure}[ht]
    \centering
    \includegraphics[width=0.48\textwidth]{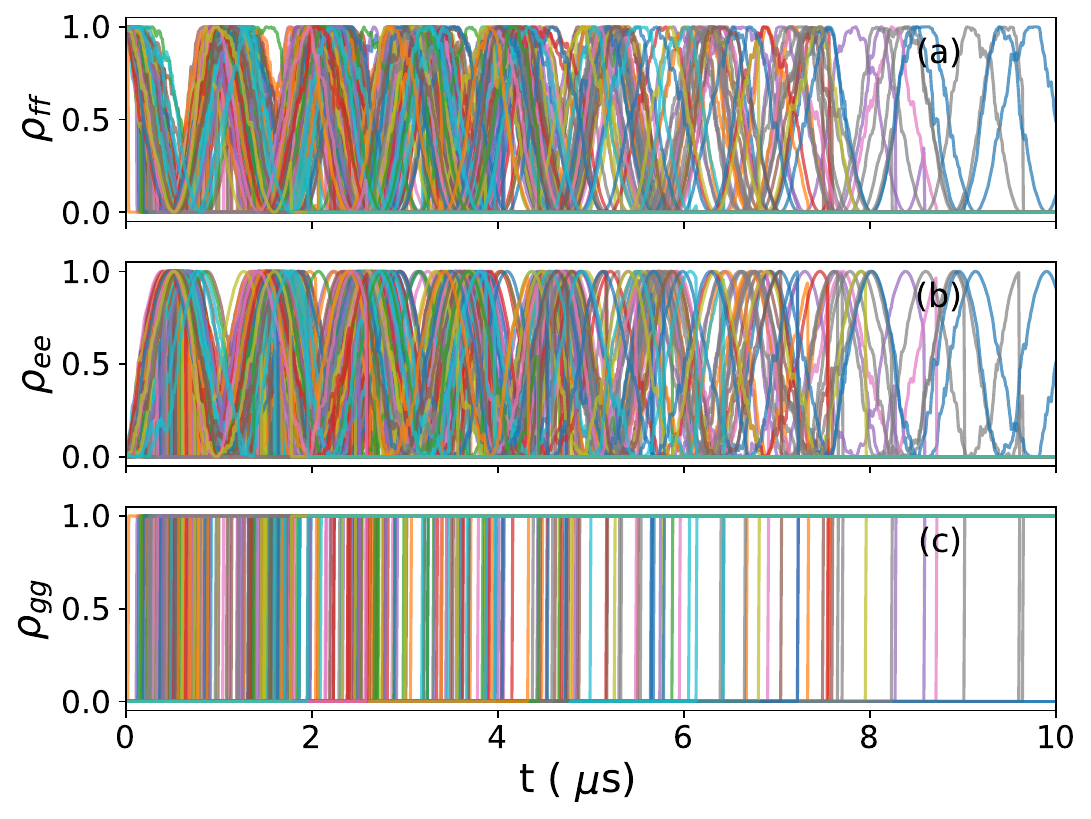}
    \includegraphics[width=0.45\textwidth]{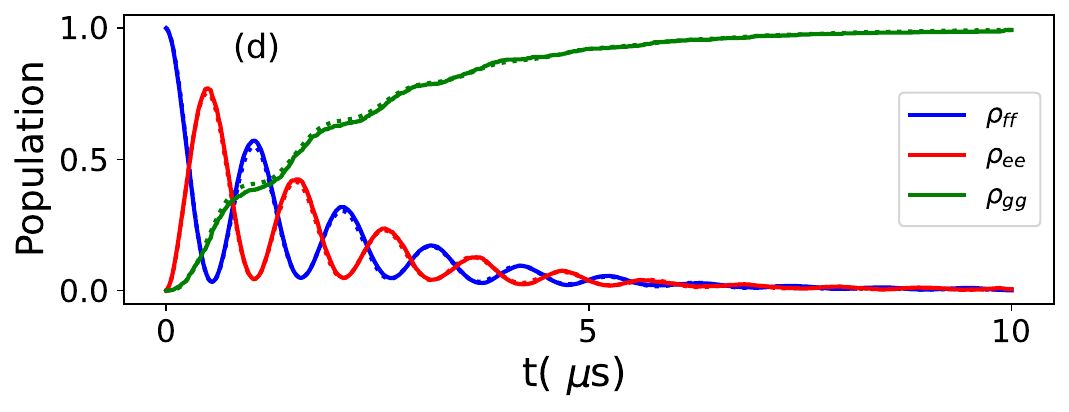}
    \caption{(a)–(c) State-update trajectories obtained from Eq.~\eqref{eq:state_uodate}. 
    The trajectories are shown in terms of the state populations 
    $\rho_{ff}$, $\rho_{ee}$, and $\rho_{gg}$, corresponding to the second excited, 
    first excited, and ground states, respectively. These populations are 
    the diagonal elements of the density matrix. 
    (b) The trajectory ensemble-averaged values of the populations are shown as solid lines. 
    This ensemble average is compared with the average dynamics of the full 
    three-level system governed by the Lindblad master equation, shown as dotted lines. The two averages match excellently.
      Parameters used: $\gamma_{g}/2\pi = 1~\text{MHz}$, $\gamma_{e}/2\pi = 0.2~\text{MHz}$, and $\omega/2\pi = 3~\text{MHz}$, with the number of trajectories $= 10^{3}$.}
    \label{fig:three_level_full}
\end{figure}

A single measurement trajectory can be obtained by updating the state at every interval $dt$ up to some total measurement time $T$. 
Each trajectory corresponds to a pure state. 
An example of such trajectories is shown in Fig.\ref{fig:three_level_full}. 
We consider the $\ket{f}$ and $\ket{e}$ states to be coherently driven such that $U = \exp(-i \omega (|f><e| + |e><f|)dt)$. 
As seen from the figure, when no jump from $\ket{e}$ to $\ket{g}$ occurs, the system stochastically oscillates between the $\ket{f}$ and $\ket{e}$ states, and as soon as a jump from $\ket{e}$ to $\ket{g}$ is registered, the trajectory collapses into the ground state and remains there until the end of the measurement time.
By averaging over many such trajectories, an ensemble-averaged dynamics of the system can be obtained, which corresponds to a mixed state. 

To validate the hybrid detector protocol, we compare the average dynamics obtained from the trajectory ensemble with those obtained from the Lindblad master equation. 
\begin{equation}
    \frac{d\bar\rho}{dt} = -i [H, \bar\rho] + \gamma_g \mathcal{D}[\ket{g}\bra{e}] \bar\rho + \gamma_e \mathcal{D}[\ket{e}\bra{f}] \bar\rho,
    \label{eq:average_master_three_level}
\end{equation}
Here, we have taken $ \bar\rho(t)$ to be the Liouvillian average.
We plot the average dynamics of second excited state probability $\rho_{ff}(t)$, the first excited state probability $\rho_{ee}(t)$, and the ground state probability $\rho_{gg}(t)$ from both approaches in Fig.\ref{fig:three_level_full}(d), where the probabilities match excellently.

In the experiment Ref.~\cite{naghiloo2019,Chen2021}, the measurement of the $\ket{e}$ and $\ket{f}$ postselected manifold is realized by performing a homodyne measurement of the signal emitted from the three-level system using a Josephson parametric amplifier. The homodyne measurement detects the system-state-dependent shift of the cavity frequency and thereby determines the state of the system. 
From this measurement, only the outcomes that remain in the $\ket{e}$ and $\ket{f}$ manifold are considered for analysis. 
The two measurement schemes discussed above illustrate different unraveling models of the measurement trajectories of the three-level system. These unravelings yield the same average dynamics as obtained in a typical experimental measurement. When quantum jumps to the ground state are postselected and removed from the analysis, the resulting dynamics correspond to a non-Hermitian qubit.

\section{Non-Hermitian Qubit}
\label{Non Hermitian Qubit}
A non-Hermitian qubit is a two-level system that interacts with a lossy channel, a gain channel, or both. The system evolves under an effective non-Hermitian Hamiltonian that captures the influence of these channels. Such a non-Hermitian qubit can be modeled using the measurement schemes of a two-level or a three-level system undergoing spontaneous emission, as discussed in the previous section.

In the two-level system, this is achieved by postselecting trajectories in which the quantum jump from the excited state to the ground state occurs. When the jump is included, the exceptional point (EP) is observable only during the transient stage of the evolution.
In the three-level system, the non-Hermitian qubit is realized by postselecting detector trajectories in which no jumps from $\ket{f}\rightarrow\ket{e}$ and $\ket{e}\rightarrow\ket{g}$ occur. In this case, the system exhibits \(\mathcal{PT}\) symmetry, since the coherent evolution between $\ket{f}$ and $\ket{e}$ is governed by the difference in the eigenvalues of the Liouvillian that describes the postselected dynamics confined to the $\ket{f}$ and $\ket{e}$ manifold.
If the $\ket{f}\rightarrow\ket{e}$ jump is not postselected, homodyne detection of the spontaneous decay $\ket{f}\rightarrow\ket{e}$ can be performed. By averaging over the resulting measurement trajectories, one can model a non-Hermitian fluorescent qubit formed within the $\ket{f}$ and $\ket{e}$ manifold. The average dynamics of such a non-Hermitian qubit—particularly near the exceptional point, where eigenstates coalesce—can be analyzed using the Liouvillian superoperator formalism. Later, we compare the dynamics obtained from this approach with those derived from the postselected ensemble-averaged trajectory method.

\begin{figure}[ht]
    \centering
    \includegraphics[width=0.35\textwidth]{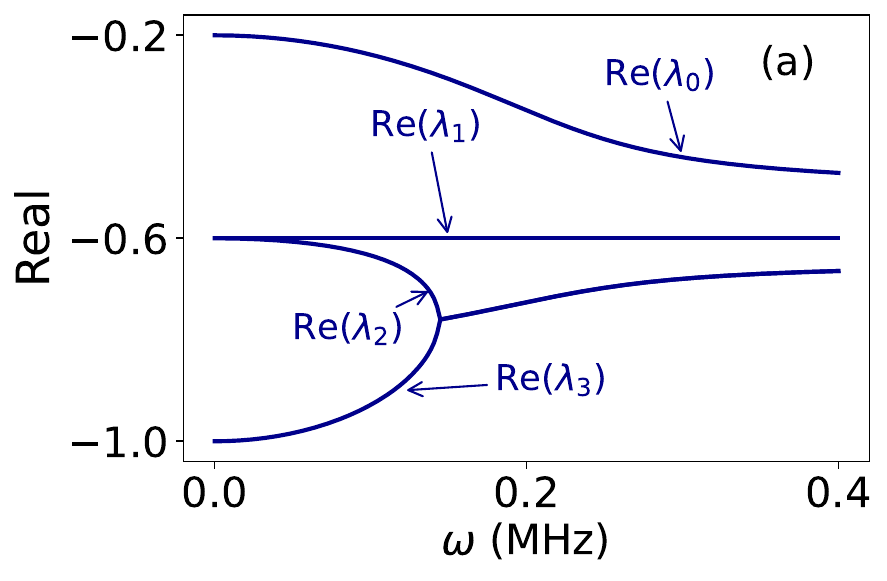}
    \includegraphics[width=0.35\textwidth]{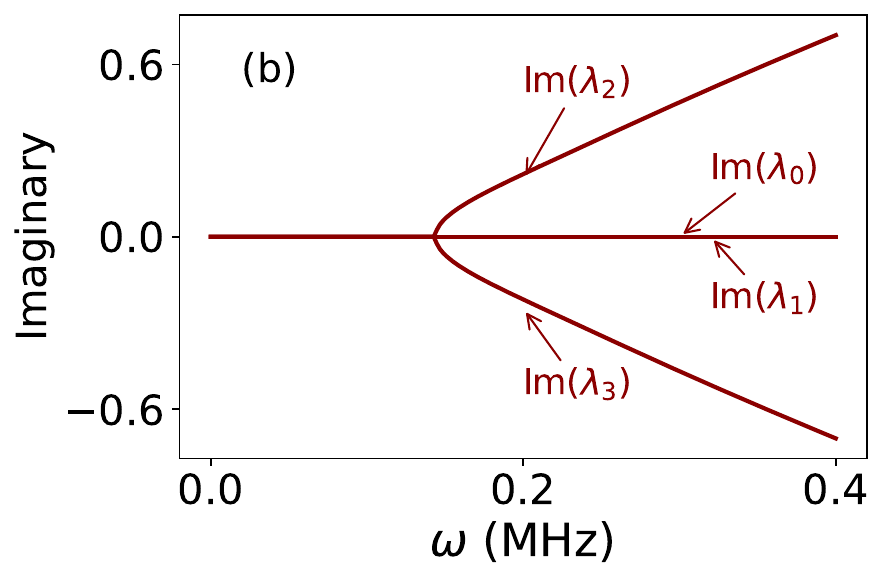}
    \caption{Eigenvalue spectrum of the Liouvillian, with the real part shown in (a) 
    and the imaginary part shown in (b). Parameters used: $\gamma_{g}/2\pi = 1~\text{MHz}$, $\gamma_{e}/2\pi = 0.2~\text{MHz}$. The system exhibits a second-order EP, which arises from the presence of the $\ket f \rightarrow \ket e$ quantum jump. If this jump is postselected, the system instead supports a third-order EP and displays $\mathcal{PT}$ symmetry.}
    \label{fig:L_spec}
\end{figure}
As shown in the Appendix \ref{appendix B}, the differential equation for the vectorized density matrix
\(
\bar\rho = (\bar\rho_{ee}, \bar\rho_{ef}, \bar\rho_{fe}, \bar\rho_{ff})^T
\)
of the $\ket{f}$ and $\ket{e}$ manifold can be written in the form
\(
\frac{d\bar\rho}{dt} = \mathcal{L} \, \bar\rho,
\)
where the Liouvillian operator $\mathcal{L}$ is given by
\begin{equation}
\label{eq:Liouvillian_matrix}
\mathcal{L} =
\begin{bmatrix}
-\gamma_e & i \omega & -i \omega & \gamma_f\\
i \omega & -(\gamma_e + \gamma_g)/2 & 0 & -i \omega \\
-i \omega & 0 & -(\gamma_e + \gamma_f)/2 & i \omega \\
0 & -i \omega & i \omega & -\gamma_f 
\end{bmatrix}.
\end{equation}
This Liouvillian is obtained for the case where the $\ket{e} \to \ket{g}$ jump is postselected while the $\ket{f} \to \ket{e}$ jump is retained. 
It has four eigenvectors; therefore, the density operator can be expressed in terms of its eigenvalue decomposition as
\begin{equation}
\bar\rho(t) = \sum_{k=1}^{4} C_k \, e^{\lambda_k t} \, R_k,
\end{equation}
where $\lambda_k$ and $R_k$ are the eigenvalues and right eigenvectors of $\mathcal{L}$, and $C_k = Tr[L_k \bar\rho(0)]$ are determined from the initial condition $\bar\rho(0)$. The left and right eigenvectors satisfy the biorthonormal condition \(\mathrm{Tr}\big[L_j \, R_k \big] = \delta_{jk}.\) 
The real parts of the eigenvalues are always negative \cite{Ming2019}, and hence they lead to relaxation of any observable toward the steady state. 
The imaginary parts, on the other hand, lead to oscillations in the observables.
The spectrum of eigenvalues as a function of the drive strength $\omega$ is shown in Fig.\ref{fig:L_spec}. As seen in the figure, we observe a second-order EP. 
On the right side of the EP, the imaginary parts are nonzero, resulting in oscillatory behavior of observables. 
Simultaneously, the real parts are negative and nonzero, which leads to decoherence at a rate proportional to the difference between the eigenvalues \cite{naghiloo2019}. 
On the left of the EP, there are no oscillations, only decay. Therefore, the EP separates the overdamped and underdamped oscillatory regimes of the qubit.  
This transition is observable only during the transient period. However, if we also postselect the 
$\ket{f} \to \ket{e}$ jump, a \(\mathcal{PT}\)-symmetric phase emerges in the steady state.

\section{Comparing Average Dynamics of Liouvillian and Trajectory Approaches}
\label{Comparing Dynamics}

\begin{figure}[ht]
    \centering
    \includegraphics[width=0.23\textwidth]{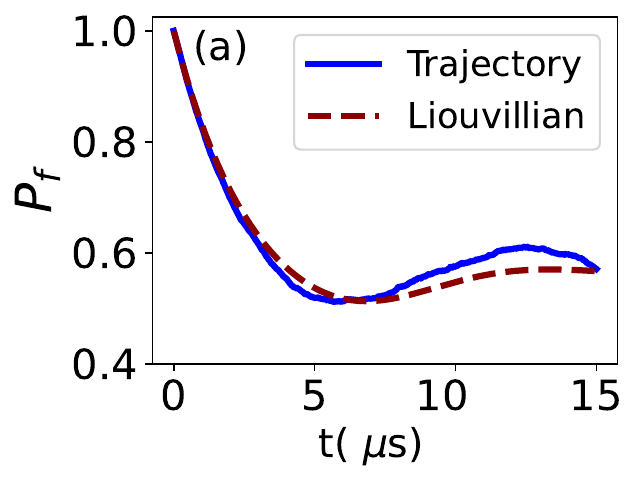}
    \includegraphics[width=0.23\textwidth]{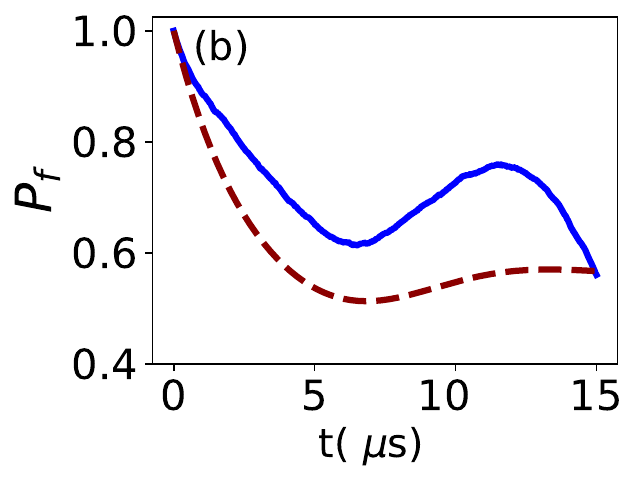}
    \includegraphics[width=0.23\textwidth]{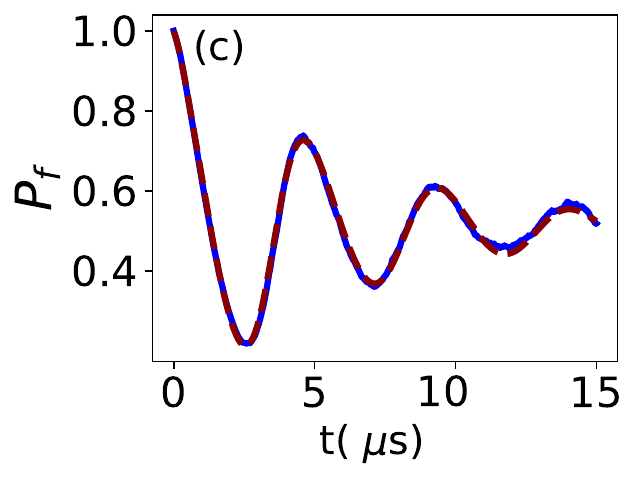}
    \includegraphics[width=0.23\textwidth]{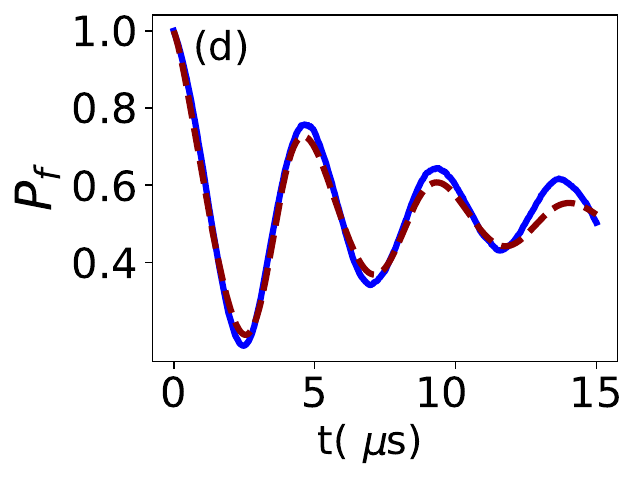}
    \includegraphics[width=0.23\textwidth]{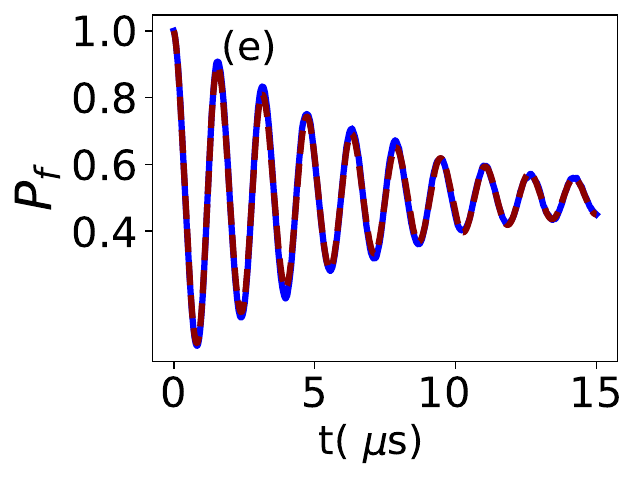}
    \includegraphics[width=0.23\textwidth]{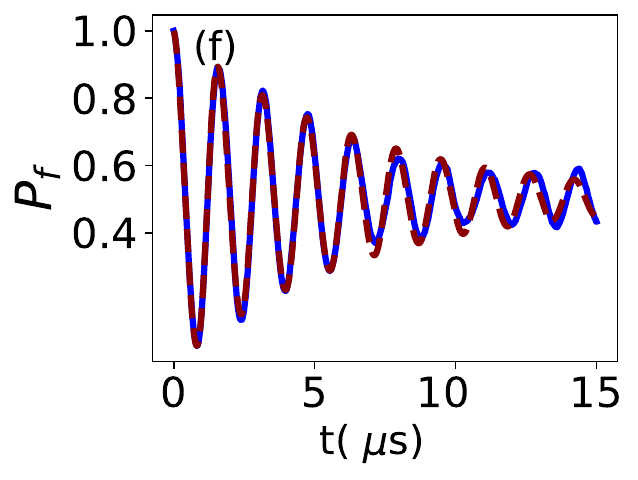}

    \caption{ The evolution of the normalized element of the density matrix, 
    i.e., the population of the $\lvert f \rangle$ state ($P_f$). 
    The solid line corresponds to the result obtained from averaging the trajectories where both jump and no-jump state updates are considered, while the dotted line corresponds to 
    the Liouvillian dynamics. 
    (a)–(b) show the case where the drive strength is close to the Liouvillian EP, 
    and (c)–(f) correspond to the case away from the EP. 
    (a), (c) and (e) correspond to driving about the $x$-axis, while (b), (d) and (f) correspond 
    to driving about the $y$-axis. 
    All plots are generated at $\theta = 0$. 
    A similar behavior is observed when $\theta = \pi/2$, except that in this case 
    the $x$-axis drive resembles (b), (d) and (f), while the $y$-axis drive resembles (a), (c) and (e). 
  Parameters used: for (a) and (b), $\omega/2\pi = 0.3~\text{MHz}$, for (c) and (d), $\omega/2\pi = 0.7~\text{MHz}$, and for (e) and (f) $\omega/2\pi = 2 ~\text{MHz}$. The common parameters are $\gamma_{e}/2\pi = 0.2~\text{MHz}$ and $\gamma_{g}/2\pi = 1~\text{MHz}$. The time step is $\text{dt}=0.01\mu s$. A total of $5 \times 10^5$ trajectories were simulated, among which approximately $600$–$1000$ trajectories were post-selected as those without quantum jumps to the ground state.}
\label{fig:xy_drive_compare_jump_uodae}
\end{figure}

In Fig.~\ref{fig:xy_drive_compare_jump_uodae}, we plot the normalized population of the $\ket{f}$ state, defined as
\(
P_f(t) = \frac{\bar\rho_{ff}(t)}{\bar\rho_{ff}(t) + \bar\rho_{ee}(t)},
\)
obtained from the Liouvillian approach, for drive strengths both close to and far from the exceptional point (EP). In the same figure, we compare this quantity with the ensemble-averaged population derived from the state-update equation, Eq.~\eqref{eq:state_uodate}, where trajectories that do not undergo a quantum jump to the $\ket{g}$ state are postselected. In this case, $P_f(t) = \langle\rho_{ff}(t)\rangle$, where $\langle\rho_{ff}(t)\rangle$ denotes the postselected trajectory average, which is already normalized. We denote $\langle .\rangle$ the trajectory ensemble average.
Note that both homodyne and jump updates, as defined in Eq.~\eqref{eq:state_uodate}, are used to generate the postselected “no-jump” trajectories. 
We do the comparison for two driving configurations: one where the qubit is driven about the $x$-axis ($U = \exp(-i \omega (|f><e| + |e><f|)dt)$) and another where it is driven about the $y$-axis ($U = \exp(-i \omega (-i|f><e| + i|e><f|)dt)$) of the Bloch sphere constituting the $\ket f$ and $\ket e$ manifold. 
The corresponding measurement record is given by $r = \sqrt{\gamma_e}(e^{i\theta}\rho_{ef} +e^{-i\theta}\rho_{fe}) + \zeta(t)$, as can be seen from Eq.\eqref{eq:gh} and detailed in Appendix~\ref{appendix C}. Depending on the measurement quadrature, we set $\theta = 0$ (measurement along the $x$-axis) or $\theta = \pi/2$ (measurement along the $y$-axis). 

The Liouvillian-averaged dynamics agree closely with the ensemble-averaged trajectory results when the system is driven along the $x$-axis, but deviations appear for driving along the $y$-axis. These deviations become significant when the drive strength is close to the exceptional point (EP), as shown in Fig.~\ref{fig:xy_drive_compare_jump_uodae} for the case where the measurement is performed along the $x$-axis, i.e., $\theta = 0$. Away from the EP, more pronounced oscillations are observed and the deviation is significantly reduced. This deviation, whose magnitude depends on the nature of the drive, can be understood from the characteristics of the trajectories in each driving case, as well as from the nonlinearity induced by postselection.

\section{Trajectory Analysis.}
\label{Trajectory Analysis}
\begin{figure}[ht]
    \centering
    \includegraphics[width=0.23\textwidth]{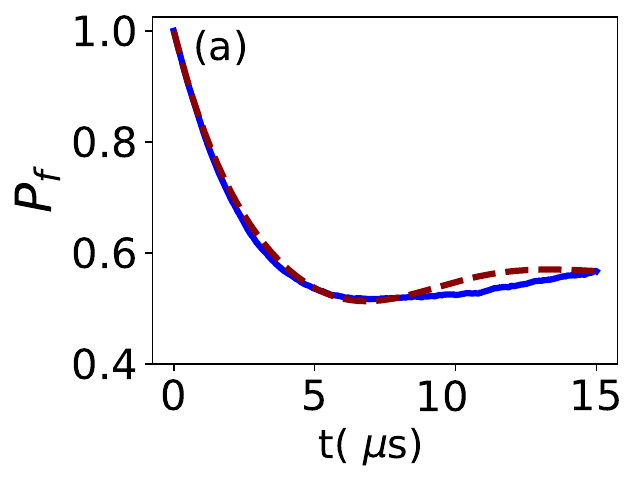}
    \includegraphics[width=0.23\textwidth]{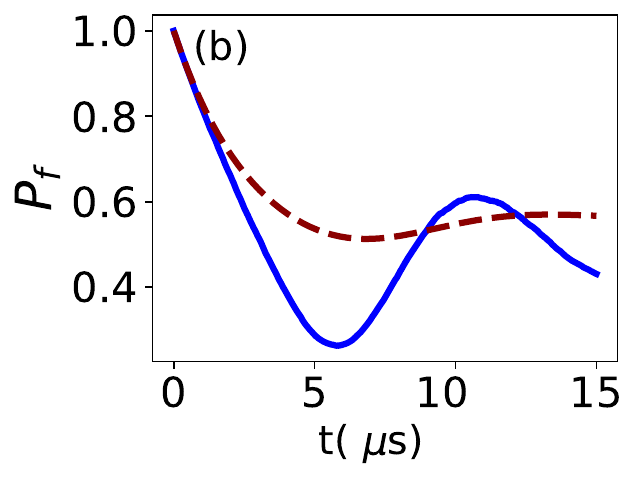}
    \includegraphics[width=0.23\textwidth]{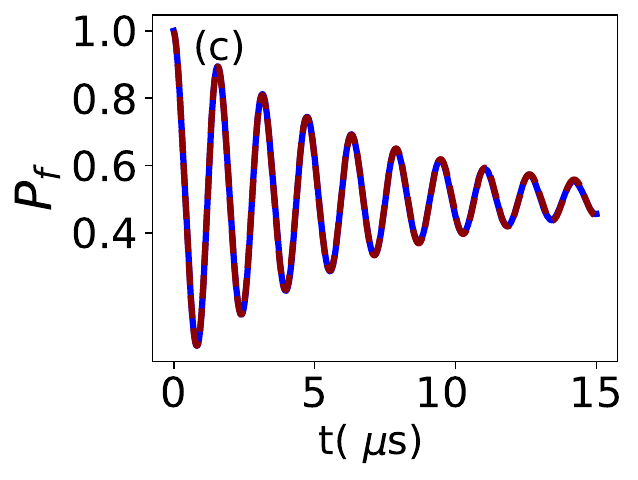}
    \includegraphics[width=0.23\textwidth]{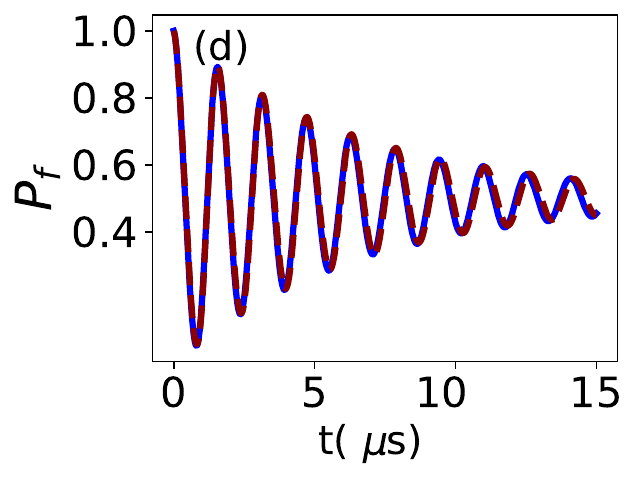}

    \caption{ The evolution of the normalized element of the density matrix, 
    i.e., the population of the $\lvert f \rangle$ state ($P_f$). 
    The solid line corresponds to the result obtained from averaging only the jump state-update trajectories, while the dotted line corresponds to 
    the Liouvillian dynamics. 
    (a)–(b) show the case where the drive strength is close to the Liouvillian EP, 
    and (c)–(d) correspond to the case away from the EP. 
    (a) and (c) correspond to driving about the $x$-axis, while (b) and (d) correspond 
    to driving about the $y$-axis. 
    All plots are generated at $\theta = 0$. 
    A similar behavior is observed when $\theta = \pi/2$, except that in this case 
    the $x$-axis drive resembles (b), while the $y$-axis drive resembles (a).
  Parameters used: for (a) and (b), $\omega/2\pi = 0.3~\text{MHz}$, and for (c) and (d), $\omega/2\pi = 2~\text{MHz}$. The common parameters are $\gamma_{e}/2\pi = 0.2~\text{MHz}$ and $\gamma_{g}/2\pi = 1~\text{MHz}$. Simulated trajectory is $5000$}
\label{fig:xy_drive_compare}
\end{figure}

\begin{figure}[t]
    \centering
    \includegraphics[width=0.23\textwidth]{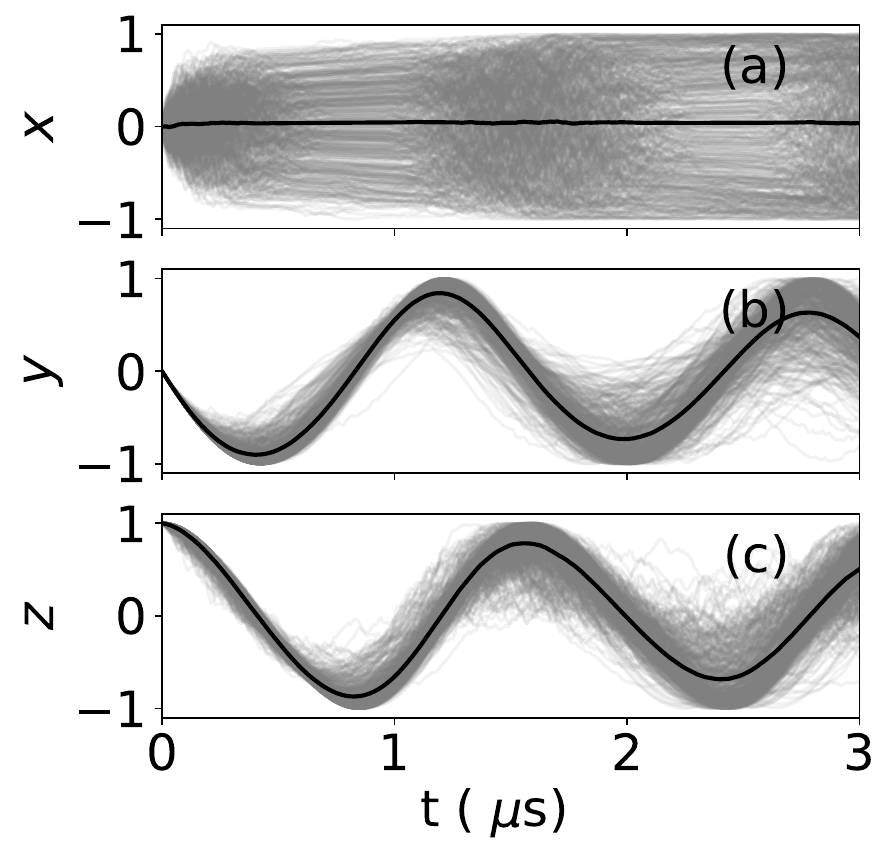}
    \includegraphics[width=0.23\textwidth]{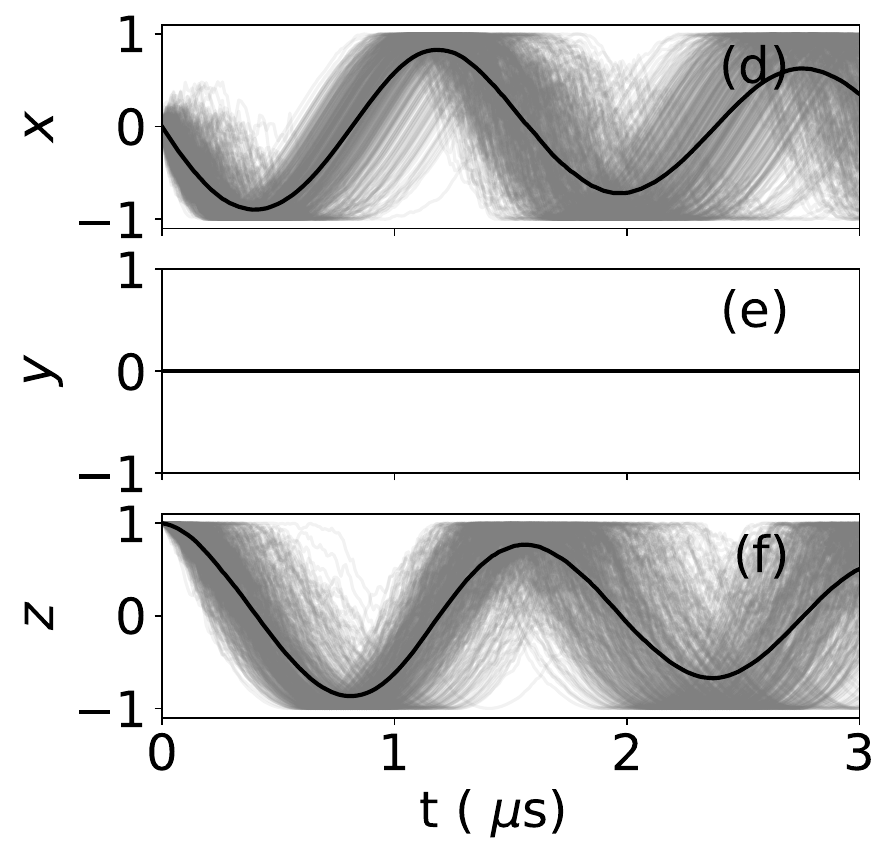}
    \caption{ Trajectory evolution of the Bloch vector components $x$, $y$, and $z$ of the 
    non-Hermitian qubit at $\theta = 0$. 
    (a)–(c) The qubit is driven about the $x$-axis. 
    The mean of the trajectories is shown as a solid black curve. 
    The qubit, initialized in the excited state, undergoes unitary oscillations about the $x$-axis. 
    This evolution drives the $y$ and $z$ components, leading to oscillations. 
    (d)–(f) The qubit is driven about the $y$-axis. 
    Here the $x$ and $z$ components oscillate unitarily, while the $y$ component remains zero. 
    This is because the qubit starts in the excited state 
    ($z = 1$, $x = 0$, $y = 0$), and the backaction on the $y$ evolution 
    is independent of $z$, leaving $y$ unchanged.  
    For $\theta = \pi/2$ measurements, the $y$-axis drive resembles the dynamics 
    in (a)–(c), while the $x$-axis drive resembles those in (d)–(f); i.e., the roles are reversed.
    The common parameters used are $\gamma_{g}/2\pi = 1~\text{MHz}$ and $\gamma_{e}/2\pi = 0.2~\text{MHz}$. 
    The drive frequency is set to $\omega/2\pi = 2~\text{MHz}$.
    }
    \label{fig:xyz_xy_drive}
\end{figure}

\begin{figure*}[ht]
    \centering
    \includegraphics[width=1\textwidth]{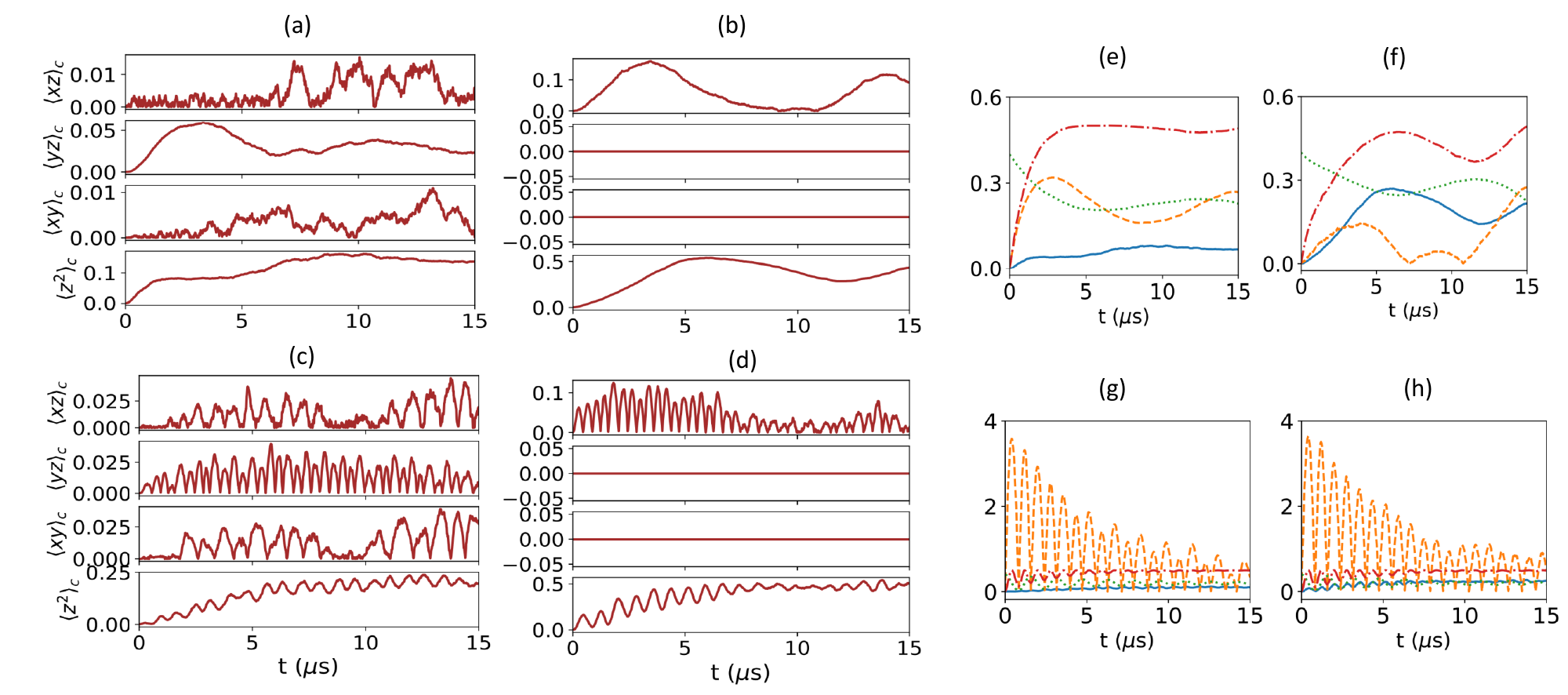}

    \caption{Comparison of covariances and contributions to the drift term in the $z$-dynamics obtained from state update Eq.~\eqref{eq:state_uodate}.
    Panels (a) and (b) show the covariances when the drive is along the $x$- and $y$-axes of the Bloch vector, respectively, at $\omega/2\pi = 0.3~\text{MHz}$. Panels (c) and (d) show the corresponding covariances for $\omega/2\pi = 2~\text{MHz}$. The comparison of the drift terms in the $z$-dynamics with the deviation $\Delta$ is shown in (e) and (g) for the $x$-drive, and in (f) and (h) for the $y$-drive. The blue solid line represents the deviation $|\Delta|$; the orange dashed line represents the magnitude of the unitary term $2\omega \langle x\rangle$ (for the $y$-drive) or $2\omega \langle y\rangle$ (for the $x$-drive); the green dotted line represents $\gamma_e(1+\langle z\rangle)$; and the red dash--dotted line represents $\gamma_g(1-\langle z\rangle^2)/2$. Panels (e) and (f) correspond to $\omega/2\pi = 0.3~\text{MHz}$, while (g) and (h) correspond to $\omega/2\pi = 2~\text{MHz}$. Other parameters are the same as in the previous plots.
    As discussed in the main text, for the $y$-drive the covariances are distributed only within the $x$–$z$ quadratures, whereas for the $x$-drive they are spread across all quadratures. Furthermore, the smaller fluctuations in the $x$-drive lead to a reduced variance contribution to the drift term compared to the $y$-drive.}
\label{fig:VarCov}
\end{figure*}

To understand the deviation between the trajectory average and the Liouvillian average, we write down the stochastic differential equation governing the state update. In terms of the Bloch vector components corresponding to the $\ket{f}$ and $\ket{e}$ manifold, the Stratonovich form of the state update for the no-jump case, $\rho_H(t+dt)$, given in Eq.~\ref{eq:state_uodate} at $\theta = 0$ (see Appendix~\ref{appendix B}), is as follows.
For the x-axis drive:
\begin{equation}
\begin{aligned}
    \dot{x} &= \frac{\gamma}{2} x z + r \sqrt{\gamma_e} (1 + z - x^2), \\
    \dot{y} &= \frac{\gamma}{2} y z - 2\omega z - r \sqrt{\gamma_e} x y, \\
    \dot{z} &= \frac{\gamma}{2} (z^2 - 1) + 2\omega y - r \sqrt{\gamma_e} x (z + 1),
    \label{eq:trajectory_eq_xdrive}
\end{aligned}
\end{equation}

For the y-axis drive:
\begin{equation}
    \begin{aligned}
        \dot{x} &= 2\omega z + \frac{\gamma}{2} x z + r \sqrt{\gamma_e} (1 + z - x^2), \\
        \dot{y} &= \frac{\gamma}{2} y z - r \sqrt{\gamma_e} x y, \\
        \dot{z} &= \frac{\gamma}{2} (z^2 - 1) - 2\omega x - r \sqrt{\gamma_e} x (z + 1),
        \label{eq:trajectory_eq_ydrive}
    \end{aligned}
\end{equation}

where $\gamma = \gamma_e - \gamma_g$ and the measurement record is 
\( r = \sqrt{\gamma_e}\, x + \zeta(t) \),
with $\zeta(t)$ being Gaussian white noise of zero mean and variance $1/dt$.
The Bloch vectors are given by $q = \mathrm{Tr}(\sigma_q \, \rho), \quad q \in \{x,y,z\}$.
The non-Hermitian decay $\gamma_g$ appears only in the drift term of the equations, while measurement backaction is determined by the characteristic measurement time $T_e = 1/\gamma_e$. 

The above equation can also be derived using the Itô stochastic master equation, where the state of the three-level system under the hybrid detection protocol is updated at each measurement interval $dt$ as
 \cite{steck2007quantum}
\begin{align}
    d\rho =\; & -i [H, \rho] \, dt - \frac{\gamma_g}{2} \{ \ket{e}\bra{e}, \rho \} \, dt + \gamma_g \langle \ket{e}\bra{e} \rangle \rho \, dt \nonumber \\
    & + \left( \frac{\ket{g}\bra{e} \rho \ket{e}\bra{g}}{\langle \ket{e}\bra{e} \rangle} - \rho \right) dN + \gamma_e \mathcal{D}[\ket{e}\bra{f}] \rho \, dt \nonumber \\
    & + \sqrt{\gamma_e} \, \mathcal{H}[\ket{e}\bra{f} e^{i\theta}] \rho \, dW,
    \label{eq:stochastic_master_Ito_full}
\end{align}
where, $H$ is the unitary Hamiltonian of the system and 
\begin{align}
    \mathcal{H}[\ket{e}\bra{f} e^{i\theta}] \rho =\; & e^{i\theta} \ket{e}\bra{f} \rho + e^{-i\theta} \rho \ket{f}\bra{e} \nonumber \\
    & - \langle e^{i\theta} \ket{e}\bra{f} + e^{-i\theta} \ket{f}\bra{e} \rangle \rho,
    \label{eq:measurement_superoperator_Ito}
\end{align}
with $dW$ being a Wiener process having zero mean, $\langle dW(t) \rangle = 0$, and $dW(t)^2 = dt$. 
The term $dN$ defines the counting process where, during the interval $dt$, it takes the value unity with probability $\gamma_g \langle \ket{e}\bra{e} \rangle dt$ and zero otherwise.
The above equation is a stochastic master equation in Itô form. 
To obtain the ensemble-averaged dynamics of the system, we take the average $\langle d\rho \rangle$, which yields the Lindblad master equation for a three-level system Eq.\eqref{eq:average_master_three_level}.

The Itô form of the stochastic master equation for the three-level system, where jumps from the $\ket{e}$ to $\ket{g}$ state are removed using post-selection, is given from Eq.~\eqref{eq:stochastic_master_Ito_full} as
\begin{align}
    d\rho =\;& -i [H, \rho] \, dt - \frac{\gamma_g}{2} \{ \ket{e}\bra{e}, \rho \} \, dt + \gamma_e \mathcal{D}[\ket{e}\bra{f}] \rho \, dt \nonumber \\
    & + \gamma_g \langle \ket{e}\bra{e} \rangle \rho \, dt + \sqrt{\gamma_e} \, \mathcal{H}[\ket{e}\bra{f} e^{i\theta}] \rho \, dW,
    \label{eq:stochastic_master_post_select_normalized}
\end{align}
In term of the Bloch components, the Itô form for $\theta =0 $ is given by
\begin{eqnarray}
    dx &=& \left( -\frac{\gamma_e}{2} x - \frac{\gamma_g}{2} x z \right) dt + \sqrt{\gamma_e} (1 + z - x^2) dW, \nonumber \\
    dy &=& \left( -2 \omega z - \frac{\gamma_e}{2} y - \frac{\gamma_g}{2} y z \right) dt + \sqrt{\gamma_e} (-x y) dW, \nonumber \\
    dz &=& \left( 2 \omega y - \gamma_e (1 + z) + \frac{\gamma_g}{2} (1 - z^2) \right) dt  \nonumber \\
       && \quad + \sqrt{\gamma_e} (-x (1 + z)) dW,
    \label{eq:Ito_form_x_drive}
\end{eqnarray}
for the $x$-axis drive, and
\begin{eqnarray}
    dx &=& \left( 2 \omega z - \frac{\gamma_e}{2} x - \frac{\gamma_g}{2} x z \right) dt + \sqrt{\gamma_e} (1 + z - x^2) dW, \nonumber \\
    dy &=& \left( -\frac{\gamma_e}{2} y - \frac{\gamma_g}{2} y z \right) dt + \sqrt{\gamma_e} (-x y) dW, \nonumber \\
    dz &=& \left( -2 \omega x - \gamma_e (1 + z) + \frac{\gamma_g}{2} (1 - z^2) \right) dt  \nonumber \\
       && \quad + \sqrt{\gamma_e} (-x (1 + z)) dW,
    \label{eq:Ito_form_y_drive}
\end{eqnarray}
for the $y$-axis drive.
The Itô forms given in Eqs.~\eqref{eq:Ito_form_x_drive} and \eqref{eq:Ito_form_y_drive} are equivalent to the Stratonovich forms given in Eqs.~\eqref{eq:trajectory_eq_xdrive} and \eqref{eq:trajectory_eq_ydrive}, respectively; they yield the same dynamics.

The trajectories obtained from the no-jump update equations (Eqs.~\eqref{eq:Ito_form_x_drive}, \eqref{eq:Ito_form_y_drive}, \eqref{eq:trajectory_eq_xdrive}, and \eqref{eq:trajectory_eq_ydrive}) also include those that could have undergone a jump to the $\ket{g}$ manifold. To accurately capture the dynamics of the non-Hermitian qubit, both the no-jump update $\rho_H(t+dt)$ and the jump update $\rho_J(t+dt)$ must be included. After applying both updates, we postselect the trajectories that have not decayed to the ground-state manifold and perform ensemble averaging over this subset.
The ensemble-averaged dynamics that include both the jump and no-jump updates are shown in Fig.~\ref{fig:xy_drive_compare_jump_uodae}, while the dynamics obtained using only the no-jump update are shown in Fig.~\ref{fig:xy_drive_compare}. The trajectories corresponding to the state updates derived from the no-jump stochastic differential equations, in both the Itô and Stratonovich forms, are shown in Fig.~\ref{fig:xyz_xy_drive}. 
Note that the jump-included postselected trajectories closely resemble the corresponding no-jump trajectories shown in Fig.~\ref{fig:xyz_xy_drive}; differences arise only after ensemble averaging, due to the postselection of jumps, as shown in Fig.~\ref{fig:xy_drive_compare}. Consequently, the essential features of the jump-included postselected trajectory can be inferred directly from the no-jump stochastic equations.

The deviation between the trajectory ensemble average and the Liouvillian average can be straightforwardly understood from the normalized Itô stochastic differential equation in the no-jump case. By taking the average of Eq.~\eqref{eq:stochastic_master_post_select_normalized} or of Eqs.~\eqref{eq:Ito_form_x_drive} and \eqref{eq:Ito_form_y_drive}, we obtain normalized populations similar to those obtained from Eq.~\eqref{eq:Liouvillian_matrix}, i.e., the Liouvillian dynamics, provided that $\langle AB\rangle = \langle A\rangle \langle B\rangle$, where $A,B = x,y,z$. However, when this average is evaluated directly from the ensemble of stochastic trajectories, the resulting dynamics do not, in general, coincide with the Liouvillian dynamics.

This can be explained as follows. \newline
(1) As evident from the It\^o differential equations (Eq.~\eqref{eq:stochastic_master_post_select_normalized}, Eqs.~\eqref{eq:Ito_form_x_drive} and \eqref{eq:Ito_form_y_drive}), the drift term contains nonlinear contributions arising from the normalization procedure, which is introduced to preserve the trace at each time step. The Liouvillian (ensemble-averaged) dynamics is obtained by taking the average of the It\^o equation. However, due to the presence of these nonlinear terms in the drift, this average does not, in general, coincide with the trajectory ensemble average. This discrepancy originates from the fact that $\langle AB\rangle \neq \langle A\rangle \langle B\rangle$, where $A,B = x,y,z$. Consequently, the degree of deviation between the trajectory ensemble average and the Liouvillian evolution is governed by the covariance between the Bloch components, defined as
\begin{equation}
\langle AB\rangle_c = \langle AB\rangle - \langle A\rangle \langle B\rangle .
\end{equation}
Larger covariances therefore lead to larger deviations. This behavior is illustrated in Fig.~\ref{fig:VarCov} for both the $x$- and $y$-driven cases evaluated from the state update Eq.~\eqref{eq:state_uodate}.
To make this explicit, we analyze the three contributions to the drift term in the $z$-dynamics: the unitary terms $2\omega \langle x\rangle$ (for the $\langle y\rangle$ drive) or $2\omega \langle y\rangle$ (for the $\langle x\rangle$ drive), and the two common dissipative terms, $\gamma_e(1+\langle z\rangle)$ and $\gamma_g(1-\langle z\rangle^2)/2$. The deviation arising from the nonlinear contribution is quantified by
\begin{equation}
\Delta = -\frac{\gamma_g}{2}\mathrm{Var}(z),
\end{equation}
which follows from writing $\langle z^2\rangle = \langle z\rangle^2 + \mathrm{Var}(z)$, where
\begin{equation}
\mathrm{Var}(z) = \langle z^2\rangle - \langle z\rangle^2
\end{equation}
is the variance of $z$. As shown in the Fig.~\ref{fig:VarCov}, the nonlinear deviation $\Delta$ is comparatively small for the $x$ drive but significantly larger for the $y$ drive. Near the exceptional point (EP), or for small $\omega$, this deviation becomes comparable to the unitary contribution, leading to a pronounced discrepancy between the trajectory ensemble average and the Liouvillian evolution in the $y$-driven case, as observed in Fig.~\ref{fig:VarCov}(e)-(f). In contrast, for large $\omega$ far from the EP, the unitary drift term $2\omega x$ or $2\omega y$ dominates over the nonlinear contribution, resulting in negligible deviation between the trajectory and Liouvillian averages, as observed in Fig.~\ref{fig:VarCov}(g)-(h).

(2) The larger deviation observed in the \emph{y}-drive compared to the \emph{x}-drive can be understood from the structure of the stochastic trajectory equations.
As is evident from the equations of motion, in the \emph{y}-drive case the Bloch components $x$ and $z$ evolve stochastically independently of the $y$ component, whereas in the \emph{x}-drive case all three Bloch components $x$, $y$, and $z$ evolve in a mutually coupled manner. Since the continuous measurement is performed along the $x$ axis, measurement-induced noise enters directly into the $x$ dynamics, and the noise terms in the $x$, $y$, and $z$ equations all contain backaction originating from fluctuations in $x$.

For the initial state $(x,y,z)=(0,0,1)$, measurement backaction first enters the dynamics through the $x$ component, since the continuous measurement is performed along the $x$ axis. In the subsequent time-step evolution, these measurement-induced fluctuations originating in $x$ are redistributed into the $x$, $y$, and $z$ dynamics through the nonlinear stochastic terms. This process continues throughout the time evolution.

In the y-drive configuration, the $x$ and $z$ components evolve stochastically independently of the $y$ component. As a result, backaction-generated fluctuations in the $y$ component are not fed back into the $x$ and $z$ dynamics. Consequently, the connected covariances generated by nonlinear correlations remain confined to the $(x,z)$ subspace, with nonzero contributions only to $\langle xz\rangle_c$ and $\langle z^2\rangle_c$, as shown in Fig.~\ref{fig:VarCov}(b)-(d). Moreover, the $y$ component does not acquire coherent dynamics in this configuration and remains at zero.
In contrast, in the x-drive configuration all three Bloch components $x$, $y$, and $z$ evolve in a mutually coupled manner. In this case, measurement-induced fluctuations propagated into the $y$ component are coherently rotated by the Hamiltonian and subsequently fed back into the $x$ and $z$ dynamics. As a result, nonlinear backaction effects generate finite connected covariances across all Bloch components, including $\langle xy\rangle_c$ and $\langle yz\rangle_c$, leading to a redistribution of fluctuations over the entire Bloch components, as shown in Fig.~Fig.~\ref{fig:VarCov}(a)-(c). Since the Hamiltonian drive coherently couples the $y$ and $z$ components, the $y$ component undergoes coherent evolution in the x-drive case, enabling finite backaction from $y$ into $x$ and $z$. 
As a consequence, the variance—and hence the deviation \(\Delta\)—is larger for the \(y\)-drive case than for the \(x\)-drive case. 
The larger variation of the trajectories for the $y$-drive compared to the $x$-drive can be observed in the trajectory plot shown in Fig.~\ref{fig:xyz_xy_drive}.

(3) Combining the arguments in (1) and (2), we deduce the following: The large variance in the $y$-drive trajectories, arising from the $x$-quadrature measurement of the qubit, leads to a large variation in the drift term due to the variance-dependent nonlinear term $\langle z^2 \rangle$. Consequently, the trajectory ensemble average deviates significantly from the Liouvillian average. In contrast, for the $x$-drive, this variance is smaller, and the deviation from the Liouvillian dynamics is therefore weaker.

When the measurement axis is rotated to $y$ (corresponding to $\theta=\pi/2$), the roles of the x- and y-drive configurations are interchanged. Since the continuous measurement is performed along the $y$ axis, measurement-induced noise enters directly into the $y$ dynamics, and the noise terms in the $x$, $y$, and $z$ equations all contain backaction originating from fluctuations in $y$, as can be seen from Eq.~\eqref{eq:measurement_superoperator_Ito} and \eqref{eq:trajectory_appebdix_sigmay_drive}. Consequently, in this configuration the y-drive exhibits a smaller deviation compared to the x-drive when compared with the corresponding Liouvillian dynamics.

\begin{figure}[ht]
    \centering
    \includegraphics[width=0.23\textwidth]{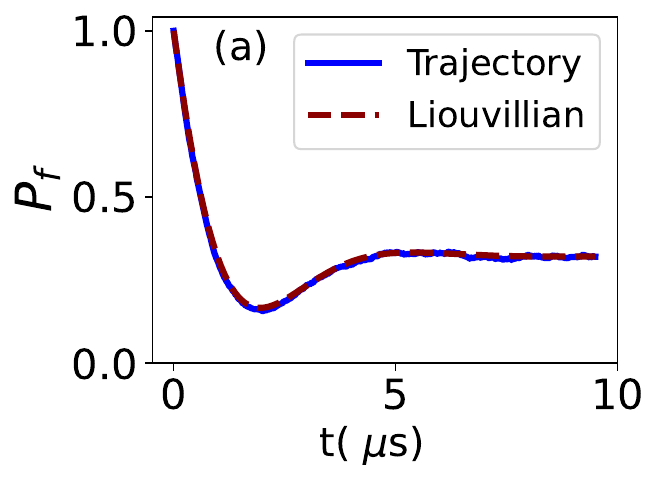}
    \includegraphics[width=0.23\textwidth]{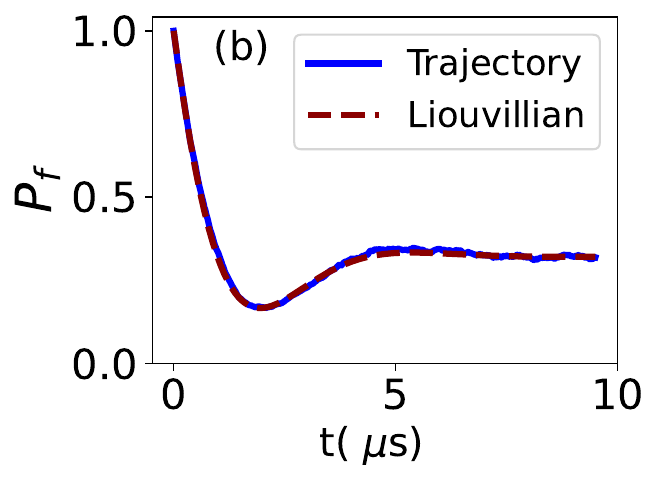}

    \caption{Comparison of the trajectory average and the Liouvillian average 
    in terms of the $\lvert f \rangle$ state population for the case when 
    the non-Hermitian loss rate is $\gamma_{g}/2\pi = 0$. 
    (a) corresponds to driving about the $x$-axis, and (b) corresponds to 
    driving about the $y$-axis. 
    Irrespective of the drive axis, both averages are identical. The parameters used are $\gamma_{e}/2\pi = 1~\text{MHz}$, $\gamma_{g}/2\pi = 0~\text{MHz}$, and $\omega/2\pi = 0.3~\text{MHz}$.}
\label{fig:xy_drive_gam_g_0}
\end{figure}

\begin{figure}[ht]
    \centering
    \includegraphics[width=0.22\textwidth]{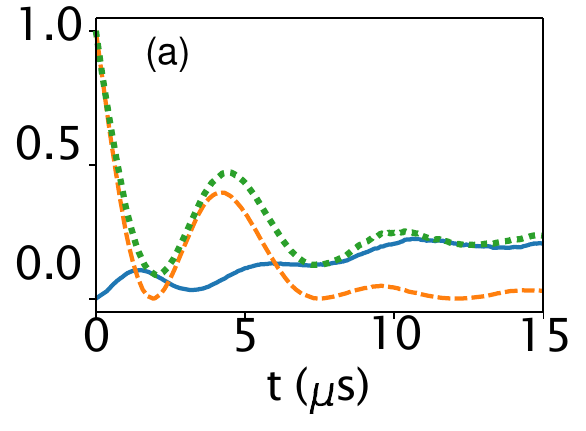}
    \includegraphics[width=0.22\textwidth]{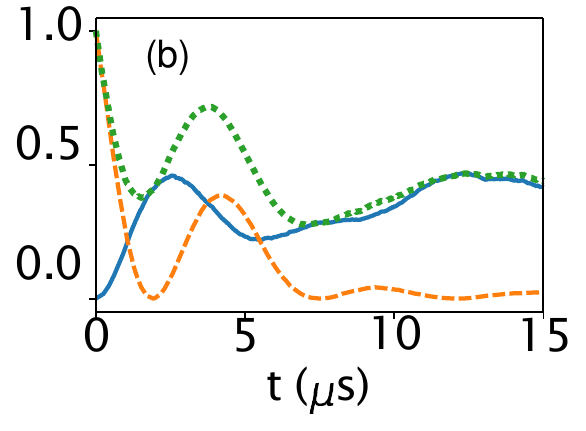}
    \includegraphics[width=0.22\textwidth]{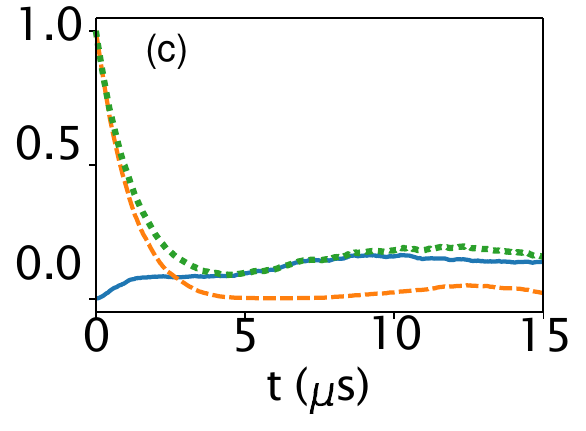}
    \includegraphics[width=0.22\textwidth]{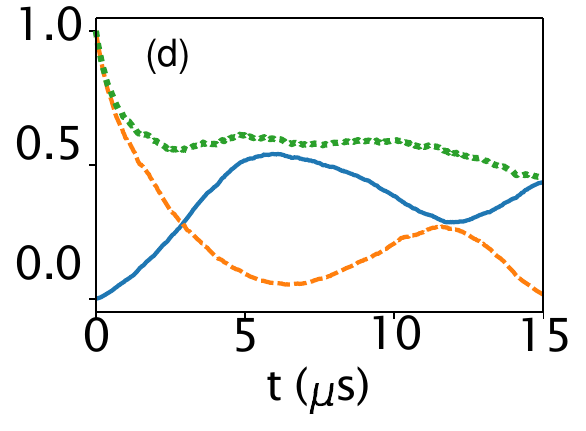}   
    \caption{Time evolution of $\mathrm{Var}(z)$ (blue solid line), $\langle z^2\rangle$ (green dotted line), and $\langle z\rangle^2$ (orange dashed line) for the $x$ drive [(a), (c)] and the $y$ drive [(b), (d)]. Panels (a) and (b) correspond to a simple two-level system, while panels (c) and (d) show the postselected non-Hermitian case. For (a) and (b), the parameters are $\gamma_{e}/2\pi = 0.2~\text{MHz}$, $\gamma_{g}/2\pi = 0$~ and $\omega/2\pi = 0.3~\text{MHz}$. For (c) and (d), $\gamma_{e}/2\pi = 0.2~\text{MHz}$, $\gamma_{g}/2\pi = 1~\text{MHz}$, and $\omega/2\pi = 0.3~\text{MHz}$.}
\label{fig:moment}
\end{figure}

In the absence of the decay $\gamma_g$, the system reduces to a simple two-level system with no non-Hermitian loss channel. In this case, there are no quantum jumps to the ground state, and the nonlinearity induced by postselection is no longer present. Hence, the trajectory ensemble average perfectly matches that of the Liouvillian approach, as shown in Fig.~\ref{fig:xy_drive_gam_g_0}. Here too, the trajectory variance for the $y$-drive is larger than for the $x$-drive when the measurement is performed in the $x$ quadrature, and conversely, it is larger for the $x$-drive than for the $y$-drive when measured in the $y$ quadrature. However, since the drift term in this case is linear, the trajectory ensemble average perfectly matches with the Liouvillian evolution.
This behavior is illustrated in Fig.~\ref{fig:moment}, where we plot $\mathrm{Var}(z)$, $\langle z^2\rangle$, and $\langle z\rangle^2$ for both the $x$ and $y$ drives. As expected, the linear contribution $\langle z\rangle^2$ is identical for the two drives, reflecting the fact that the mean-field (Liouvillian) evolution of $\langle z\rangle$ is insensitive to the choice of drive direction. In contrast, the variance $\mathrm{Var}(z)$ and the second moment $\langle z^2\rangle$ differ significantly and are larger for the $y$ drive, indicating stronger fluctuation buildup at the trajectory level.
For the postselected non-Hermitian qubit, however, even $\langle z\rangle^2$ differs between the two drives, as shown in Fig.~\ref{fig:moment}(c)-(d). This originates from the nonlinear term in the drift, which arises from state normalization. As a result, the ensemble-averaged dynamics no longer coincides with the linear Liouvillian prediction, and the mean evolution itself becomes sensitive to the drive through nonlinear corrections.

\section{Zero-Variance Case}

\begin{figure}[ht]
    \centering
    \includegraphics[width=0.23\textwidth]{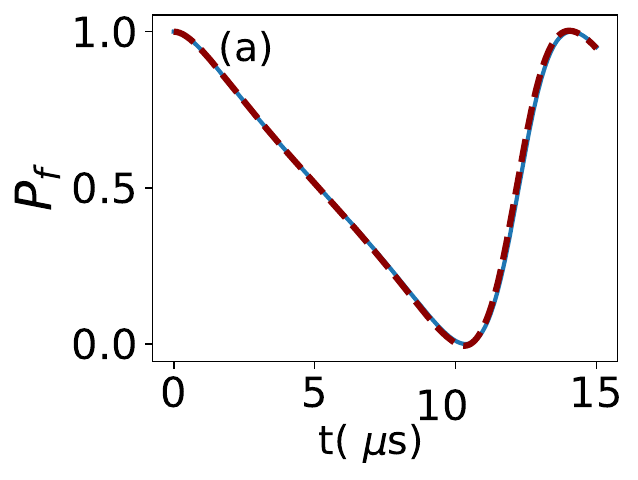}
    \includegraphics[width=0.23\textwidth]{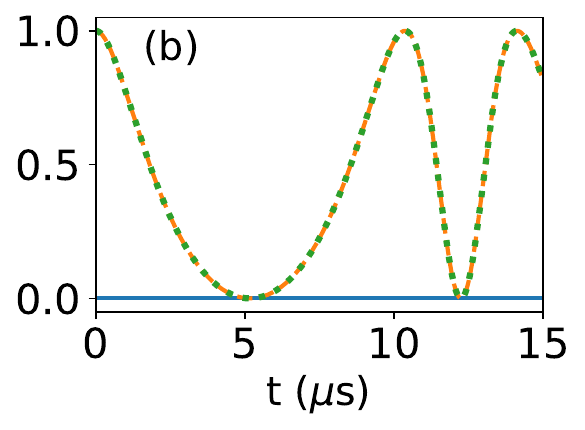}

    \caption{(a) Normalized excited-state population under the photodetection trajectory unraveling with both quantum jumps excluded. The solid line shows the result from the trajectory ensemble average, while the dotted line corresponds to the Liouvillian prediction from the normalized master equation Eq.~\eqref{eq:Photo_no_jump_at_all}.
    (b) Time evolution of $\mathrm{Var}(z)$ (blue solid), $\langle z^2\rangle$ (green dotted), and $\langle z\rangle^2$ (orange dashed). Panels (a) and (b) are identical for both $x$- and $y$-axis driving. The parameters are $\gamma_{e}/2\pi = 0.2~\mathrm{MHz}$, $\gamma_{g}/2\pi = 1~\mathrm{MHz}$, and $\omega/2\pi = 0.3~\mathrm{MHz}$.}
\label{fig:photo_no_jump}
\end{figure}

The variance in the trajectories obtained from homodyne measurement arises from the stochastic quantum jump nature of the process. If we postselect the quantum jump from the \( \ket f \to \ket e \) state in addition to the postselection of \( \ket e \to \ket g \), the variance can be removed. This can be achieved in the photodetection scheme where both jumps are detected (Fig.~\ref{fig:System}(a)). In this case, the state updates are governed by the Kraus operators \( K_0 \), \( K_{1e} \), and \( K_{1g} \). From the state updates generated by these Kraus operators, one can postselect only those trajectories in which both jumps are not detected. The ensemble average of all such trajectories is shown in Fig.~\ref{fig:photo_no_jump}.
To compare with the Liouvillian average, one can directly solve the average dynamics of the normalized master equation obtained for the postselected case, given below:
\begin{equation}
\begin{aligned}
\frac{d\bar\rho}{dt} &= -\tfrac{i}{\hbar}[H,\bar\rho]
        - \tfrac{\gamma_g}{2}\{\ket{e}\bra{e},\bar\rho\}
        - \tfrac{\gamma_e}{2}\{\ket{f}\bra{f},\bar\rho\}\\
      &\quad
        + \gamma_g \braket{\ket{e}\bra{e}} \bar\rho
        + \gamma_e \braket{\ket{f}\bra{f}} \bar\rho.
        \label{eq:Photo_no_jump_at_all}
\end{aligned}
\end{equation}

The last two terms ensure normalization and introduce nonlinearity into the equation. Comparing the population obtained from this equation, \( P_f = \bar\rho_{ff} \), with that obtained from the trajectory ensemble average, \( P_f = \langle \rho_{ff} \rangle \), we find that both agree perfectly, as shown in Fig.~\ref{fig:photo_no_jump}. Unlike the previous case, where the variance influences the nonlinear part of the drift term, here the variance is zero and the relation \( \langle AB \rangle = \langle A \rangle \langle B \rangle \) holds. This is illustrated in Fig.~\ref{fig:photo_no_jump} for the variance of \( z \), where we observe that \( \langle z^2 \rangle = \langle z \rangle^2 \).
The reason the variance vanishes is that the quantum jumps are postselected, and the fluctuations induced by them are therefore absent. As a result, all postselected trajectories coincide with the average trajectory. This holds for both \( x \)- and \( y \)-driving.

\begin{figure}[t]
    \centering
    \includegraphics[width=0.23\textwidth]{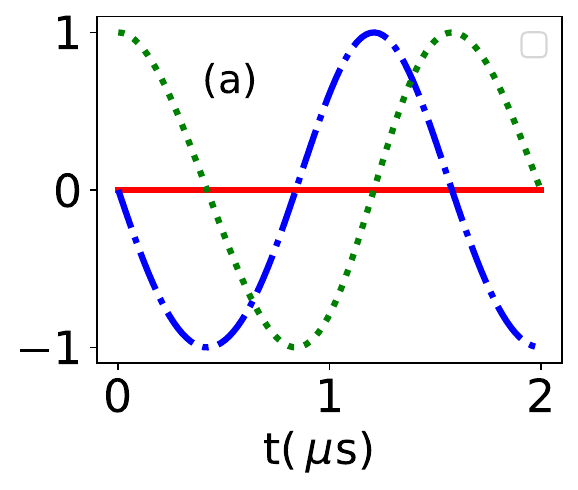}
    \includegraphics[width=0.23\textwidth]{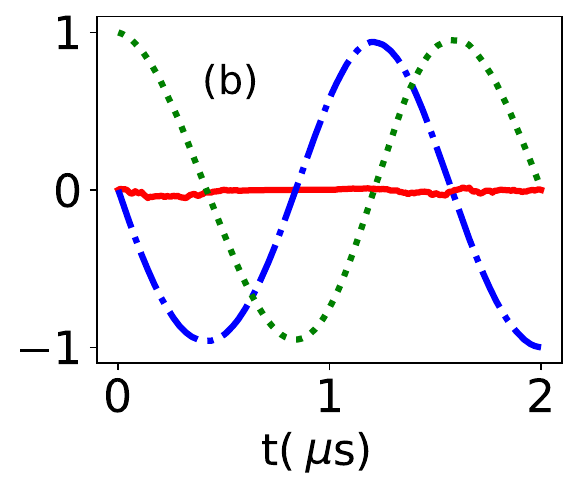}
    \caption{The average optimal path of the non-Hermitian qubit under post-selection of the final state. (a) shows the ensemble-averaged trajectory, while (b) shows the path obtained from the extremization of the action for the $x$-axis drive. The dashed red, dash-dotted blue, and dotted green curves correspond to the Bloch vector components $x$, $y$, and $z$, respectively.
    The postselected state is $q_{f} = [0,-1,0]$.
    The common parameters used are $\gamma_{g}/2\pi = 1~\text{MHz}$ and $\gamma_{e}/2\pi = 0.2~\text{MHz}$, with the drive frequency set to $\omega/2\pi = 2~\text{MHz}$. $\lambda=0.05$}
    \label{fig:x_drive_MLP}
\end{figure}

\section{Optimal path.}
\label{Optimal path}
Every trajectory has an associated action, similar to the action in the path integral formalism.  
The exponential of the action determines the probability of evolving from an initial state to a final state.  
The action for each trajectory is defined as \cite{PhysRevA.88.042110,PhysRevA.92.032125}  
\begin{equation}
S = \int_0^T dt \, \Big( -p \cdot \dot{q} + \mathcal{H}(q,p,r) \Big),
\label{eq:action_definition}
\end{equation}
where $\mathcal{H}(q,p,r)$ is the Hamiltonian (see Appendix \ref{appendix D}).  
Here, $q = [x,y,z]$ are the Bloch components, $p$ is the conjugate momentum, and $r$ is the measurement record.  

By extremizing the action, i.e., setting $\delta S = 0$, we obtain a set of constraint equations that define the optimal path of the trajectories.  
Given an initial state $q_i$ and a final state $q_f$, the optimal paths for $\sigma_x$ drive with $\theta = 0$ is shown in Fig.~\ref{fig:x_drive_MLP}.  
We also compare the optimal path with the trajectory obtained by averaging postselected trajectories.  
The trajectories are postselected by imposing the condition $\lvert q_n - q_f \rvert < \lambda$, where $\lambda$ is a tolerance parameter less than unity.  
As shown in Fig.~\ref{fig:x_drive_MLP}, the postselected average trajectory matches the optimal path obtained from extremizing the action.  

\begin{figure}[ht]
    \centering
    \begin{overpic}[width=0.34\textwidth]{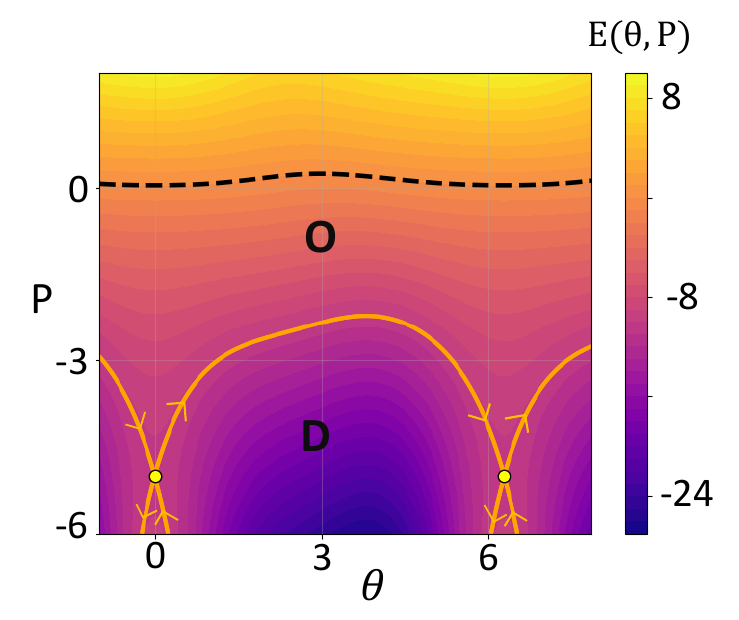}
        \put(-3,70){(a)}
    \end{overpic}


    \begin{overpic}[width=0.33\textwidth]{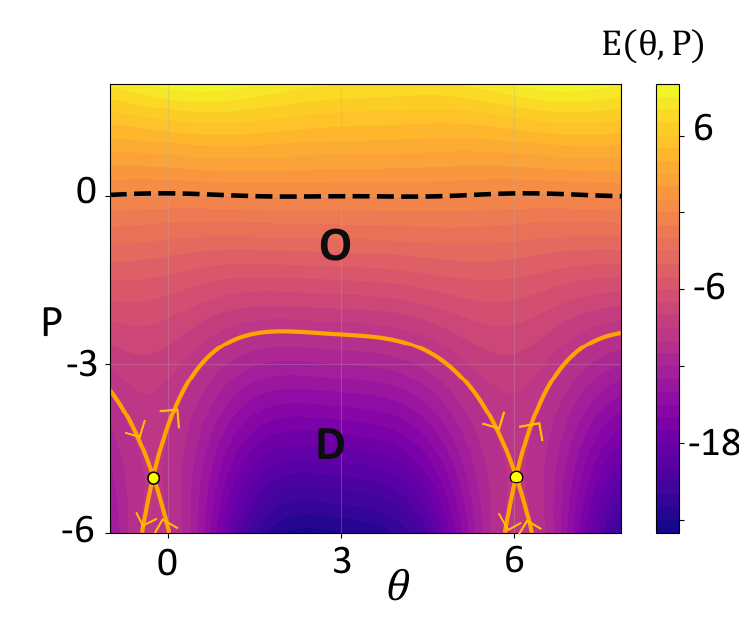}
        \put(0,70){(b)}
    \end{overpic}
    \caption{
    Phase-space plots of the qubit’s optimal paths for different values of the energy constant \(E(\theta,p)\). 
    (a) For \(\omega/2\pi = 2 ~\text{MHz}\) and \(\gamma_g/2\pi = 1\,\text{MHz}\), two saddle fixed points are observed at \((\theta,p) = (0.0,-0.65)\) and \((2\pi,-0.65)\), indicated by yellow circles. 
    In the region labeled \(\textbf{O}\), the paths oscillate between \(\theta=0\) (\(|e\rangle\)) and \(\theta=2\pi\) (\(|f\rangle\)). 
    The black dotted curve denotes a trajectory crossing the \(p=0\) line. 
    In region \(\textbf{D}\), trajectories approach \(\theta=\pi\) (\(|e\rangle\)) in the clockwise direction.
    (b) For \(\omega/2\pi = 2 ~\text{MHz}\) and \(\gamma_g/2\pi = 0~\text{MHz}\), the fixed points are shifted. This corresponds to a generic two level system in the $|f\rangle$ and $|e\rangle$ manifold.  
    Common parameters: \(\gamma_e/2\pi = 0.2\,\text{MHz}\).
    }
    \label{fig:phase_space}
\end{figure}

We can further perform a one-dimensional phase-space analysis of the qubit optimal path. 
By substituting $y = 0$, $x = \sin\theta$, and $z = \cos\theta$ into the optimal path equation for the $y$-axis drive under $y$-vector measurement of the qubit, we can confine the qubit dynamics to the $x$--$z$ plane of the Bloch sphere (see Appendix~\ref{appendix D}).  
Note that this $\theta$ is different from the phase difference $\theta$ used for measurement.
The paths corresponds to a pure state, with $\theta = 0 \rightarrow \ket{e}$ and $\theta = \pi \rightarrow \ket{f}$.  
The corresponding Hamiltonian $\mathcal{H}(\theta, p)$ (see Appendix \ref{appendix D}) is a constant of motion $E$ along the optimal path.  
Consequently, one can plot the conjugate momentum $p(\theta,E)$ to visualize the qubit optimal trajectory in phase space.  
The phase portrait for different energy constants is shown in Fig.~\ref{fig:phase_space}. 
Compared to a generic two-level system, the non-Hermitian two-level system exhibits a similar phase-space structure \cite{lewalle2020, lewallethesis}.
In the non-Hermitian case, the fixed points positions are shifted.
Of all the paths that start from a particular state $q_i$, 
the one that ends at $p_f = 0$ extremizes the action.

\section{Conclusion.}
\label{Conclusion.}
In summary, we have investigated a homodyne measurement of a two-level non-Hermitian qubit realized 
through post-selection of a three-level system. 
The post-selection procedure, which discards quantum jump to the ground-state 
manifold $\lvert g \rangle$ while retaining excitations in the 
$\lvert e \rangle$ and $\lvert f \rangle$ manifolds, effectively generates 
a non-Hermitian qubit exhibiting $\mathcal{PT}$ symmetry. 
Building on this framework, we have analyzed the system under continuous 
homodyne measurement during spontaneous emission, thereby probing the 
interplay between non-Hermitian decay (decay without quantum jump to ground state) and measurement backaction. 
Our results show that, the no-jump stochastic differential equation describing the postselected non-Hermitian qubit agree with those of the jump-updated postselected or Liouvillian evolution at drive strengths far from the Liouvillian exceptional point (EP).
The degree of deviation near the EP depends sensitively on the nature of the drive, with qualitative differences depending on whether the qubit is driven about the $x$- or $y$-axis of the Bloch sphere. 
Furthermore, by employing a path-integral formulation within the quantum 
trajectory approach, we identified the most likely measurement paths under 
post-selection. 
These findings highlight how measurement backaction and non-Hermitian 
dynamics together shape the transient behavior of open quantum systems, 
and they provide a pathway for exploring controlled dynamics near 
exceptional points.

\section*{Acknowledgement}
This work is supported by MoE, Government of India (Grant No. MoE-STARS/STARS-2/2023-0161).


\appendix
\section{Liouvilian of the Post-Selected $\ket{f}$ and $\ket{e}$ manifold qubit.}
\label{appendix B}

The Itô form of the stochastic master equation for the three-level system, where jumps from the $\ket{e}$ to $\ket{g}$ state are removed using post-selection, is given from Eq.~\eqref{eq:stochastic_master_Ito_full} as
\begin{align}
    d\rho =\; & -i [H, \rho] \, dt - \frac{\gamma_g}{2} \{ \ket{e}\bra{e}, \rho \} \, dt + \gamma_e \mathcal{D}[\ket{e}\bra{f}] \rho \, dt \nonumber \\
    & + \sqrt{\gamma_e} \, \mathcal{H}[\ket{e}\bra{f} e^{i\theta}] \rho \, dW,
    \label{eq:stochastic_master_post_select}
\end{align}
where the normalization factor (the third term in Eq.~\eqref{eq:stochastic_master_post_select}) is not taken into account.
By taking the ensemble average of Eq.~\eqref{eq:stochastic_master_post_select}, the unmonitored master equation for the fluorescent non-Hermitian qubit in the $\ket{f}$-$\ket{e}$ manifold can be written as
\begin{align}
    \frac{d\bar\rho}{dt} =\; & -i [H, \bar\rho] - \frac{\gamma_g}{2} \{ \ket{e}\bra{e}, \bar\rho \} + \gamma_e \mathcal{D}[\ket{e}\bra{f}] \bar\rho.
    \label{eq:unmonitored_master}
\end{align}
We have assumed $\langle \rho(t)\, dW(t) \rangle = 0$.
In terms of the Liouvillian operator, this can be written as
\begin{equation}
    \frac{d\bar\rho}{dt} = \mathcal{L} \, \bar\rho,
    \label{eq:Liouvilian diff}
\end{equation}
where $\bar\rho = (\bar\rho_{ff}, \bar\rho_{fe}, \bar\rho_{ef}, \bar\rho_{ee})^T$ and
\begin{align}
    \mathcal{L} =\; & -i (H \otimes I - I \otimes H^T) - \frac{\gamma_g}{2} \ket{e}\bra{e} \otimes I - \frac{\gamma_g}{2} I \otimes \ket{e}\bra{e} \nonumber \\
    & - \frac{\gamma_e}{2} \ket{f}\bra{f} \otimes I - \frac{\gamma_e}{2} I \otimes \ket{f}\bra{f} + \gamma_e \ket{e}\bra{f} \otimes \ket{e}\bra{f},
\end{align}
resulting in the matrix form given in Eq.~\eqref{eq:Liouvillian_matrix}. 
The normalized population $P_f \rightarrow \frac{\bar\rho_{ff}(t)}{\bar\rho_{ff}(t) + \bar\rho_{ee}(t)}$ obtained by solving Eq.~\ref{eq:Liouvilian diff} can also be obtained by solving the normalized It\^{o} master equation
\begin{align}
    d\bar\rho =\;& -i [H, \bar\rho] \, dt 
    - \frac{\gamma_g}{2} \{ \ket{e}\bra{e}, \bar\rho \} \, dt 
    + \gamma_e \mathcal{D}[\ket{e}\bra{f}] \bar\rho \, dt \nonumber \\
    & + \gamma_g \langle \ket{e}\bra{e} \rangle \bar\rho \, dt.
    \label{eq:Ito_post_select_normalized}
\end{align}

\section{State Update of the Post-Selected trajectories.}
\label{appendix C}

The state update of the system when no jump is registered in the detector is given by
\begin{equation}
    \rho(t + dt)|_{H} = \frac{U K_{H} \, \rho(t) \, K_{H}^\dagger U^\dagger}{\mathrm{tr} \left( U K_{H} \, \rho(t) \, K_{H}^\dagger U^\dagger \right)},
    \label{eq:state_update_post_select}
\end{equation}
where $U = \exp\left(-i \omega (|f\rangle\langle e| + |e\rangle\langle f|)dt\right)$ for the x-axis drive, and $U = \exp\left(-i \omega (-i|f\rangle\langle e| + i|e\rangle\langle f|)dt\right)$ for the y-axis drive.
By expanding the Kraus and unitary operators up to first order in $dt$, the state update can be written as
\begin{equation}
    \frac{\rho(t + dt) - \rho(t)}{dt} = i [\rho, H] + (\Sigma \rho + \rho \Sigma^\dagger) - \rho \, \mathrm{tr}(\Sigma \rho + \rho \Sigma^\dagger),
    \label{eq:rho_update}
\end{equation}
where
\begin{equation}
    \Sigma = \begin{bmatrix}
    - \frac{r^2}{4} - \frac{\gamma_e}{2} & 0 & 0 \\
    \sqrt{\gamma_e} r & - \frac{r^2}{4} - \frac{\gamma_g}{2} & 0 \\
    0 & 0 & - \frac{r^2}{4}
    \end{bmatrix}.
\end{equation}
Since the population of the system does not decay to the ground state (because no jump is registered), we consider only the $\lvert f \rangle$–$\lvert e \rangle$ manifold. 
The last term of Eq.~\eqref{eq:rho_update} can be written as
\begin{align}
\rho \, \mathrm{Tr}(\Sigma \rho + \rho \Sigma^\dagger)
&= \rho_{ff}\,(-r^2/2 - \gamma_e)  + \rho_{ee}\,(-r^2/2 - \gamma_g) \nonumber \\
&\quad + \rho_{gg}\,(-r^2/2)  + \rho_{ef}\,\sqrt{\gamma_e}\,r
+ \rho_{fe}\,\sqrt{\gamma_e}\,r.
\label{eq:trace}
\end{align}
Expanding Eq.~\eqref{eq:rho_update}, we obtain the differential equations for the $\ket{f}$--$\ket{e}$ manifold, namely $\dot{\rho}_{ff}$, $\dot{\rho}_{ee}$, $\dot{\rho}_{fe}$, and $\dot{\rho}_{ef}$. 
In these equations, the contribution from the $\ket{g}$ state appears through $\rho_{gg}$ in Eq.~\eqref{eq:trace}. 
Considering that the qubit starts in the $\ket{f}$ state, we postselect those trajectories that do not decay to the ground state $\ket{g}$. This post-selection condition effectively enforces $\rho_{gg} = 0$ throughout the evolution of the $\ket{f}$--$\ket{e}$ postselected trajectories.
Then, in terms of the Bloch vector components $x$, $y$, and $z$ used to represent a typical two-level system, the update equation in Stratonovich form can be written using Eq.~\eqref{eq:state_update_post_select} as follows:
For the x-axis drive, i.e., $H = \omega(|f\rangle\langle e| + |e\rangle\langle f|)$:
\begin{equation}
\begin{aligned}
    \dot{x} &= \frac{\gamma}{2} xz + r \sqrt{\gamma_e} \left[ (1 + z - x^2)\cos\theta + x y \sin\theta \right], \\
    \dot{y} &= \frac{\gamma}{2} y z - 2\omega z + r \sqrt{\gamma_e} \left[ (y^2 - z - 1)\sin\theta - x y \cos\theta \right], \\
    \dot{z} &= \frac{\gamma}{2}(z^2 - 1) + 2\omega y + r \sqrt{\gamma_e} (y \sin\theta - x \cos\theta)(z+1),
    \label{eq:trajectory_appebdix_sigmax_drive}
\end{aligned}
\end{equation}
For the y-axis drive, i.e., $H = \omega(-i|f\rangle\langle e| + i|e\rangle\langle f|)$:
\begin{equation}
\begin{aligned}
    \dot{x} &= 2\omega z + \frac{\gamma}{2} xz + r \sqrt{\gamma_e} \left[ (1 + z - x^2)\cos\theta + x y \sin\theta \right], \\
    \dot{y} &= \frac{\gamma}{2} y z + r \sqrt{\gamma_e} \left[ (y^2 - z - 1)\sin\theta - x y \cos\theta \right], \\
    \dot{z} &= \frac{\gamma}{2} (z^2 - 1) - 2\omega x + r \sqrt{\gamma_e} (y \sin\theta - x \cos\theta)(z+1),
    \label{eq:trajectory_appebdix_sigmay_drive}
\end{aligned}
\end{equation}
where $\gamma = \gamma_e - \gamma_g$.
The measurement record $r$ can be inferred from the probability
\(
    P = \mathrm{tr} \left( U K_h \, \rho(t) \, K_h^\dagger U^\dagger \right),
\)
which, when expanded to first order in $dt$, becomes
\begin{align}
    P ={} & \exp\left[ \frac{dt}{2} \left( (x \cos\theta - y \sin\theta)^2 \gamma_e \right. \right. \notag \\
          & \qquad \left. \left. - \frac{1+z}{2} \gamma_e - \frac{1-z}{2} \gamma_g \right) \right] \sqrt{N} \notag \\
          & \times \exp\left[ -\frac{dt}{2} \left( r - \sqrt{\gamma_e} (x \cos\theta - y \sin\theta) \right)^2 \right].
\end{align}
This shows that the probability density is Gaussian in $r$ with variance $1/dt$ and mean $\sqrt{\gamma_e}(x \cos\theta - y \sin\theta)$. Therefore, we can write
\begin{equation}
    r = \sqrt{\gamma_e} (x \cos\theta - y \sin\theta) + \zeta(t),
\end{equation}
where the noise term $\zeta(t)$ is Gaussian with zero mean and variance $1/dt$.
The two Eqs.~\eqref{eq:trajectory_appebdix_sigmax_drive} and \eqref{eq:trajectory_appebdix_sigmay_drive} can be derived from the normalized It\^{o} stochastic master equation
\begin{align}
    d\rho =\;& -i [H, \rho] \, dt - \frac{\gamma_g}{2} \{ \ket{e}\bra{e}, \rho \} \, dt + \gamma_e \mathcal{D}[\ket{e}\bra{f}] \rho \, dt \nonumber \\
    & + \gamma_g \langle \ket{e}\bra{e} \rangle \rho \, dt + \sqrt{\gamma_e} \, \mathcal{H}[\ket{e}\bra{f} e^{i\theta}] \rho \, dW,
    \label{eq:stochastic_master_post_select_normalized_appendix}
\end{align}
provided it is correctly interpreted in the Stratonovich form. The Itô form for $\theta =0 $ is given by
\begin{eqnarray}
    dx &=& \left( -\frac{\gamma_e}{2} x - \frac{\gamma_g}{2} x z \right) dt + \sqrt{\gamma_e} (1 + z - x^2) dW, \nonumber \\
    dy &=& \left( -2 \omega z - \frac{\gamma_e}{2} y - \frac{\gamma_g}{2} y z \right) dt + \sqrt{\gamma_e} (-x y) dW, \nonumber \\
    dz &=& \left( 2 \omega y - \gamma_e (1 + z) + \frac{\gamma_g}{2} (1 - z^2) \right) dt  \nonumber \\
       && \quad + \sqrt{\gamma_e} (-x (1 + z)) dW,
    \label{eq:Ito_form_x}
\end{eqnarray}
for the $x$-axis drive, and
\begin{eqnarray}
    dx &=& \left( 2 \omega z - \frac{\gamma_e}{2} x - \frac{\gamma_g}{2} x z \right) dt + \sqrt{\gamma_e} (1 + z - x^2) dW, \nonumber \\
    dy &=& \left( -\frac{\gamma_e}{2} y - \frac{\gamma_g}{2} y z \right) dt + \sqrt{\gamma_e} (-x y) dW, \nonumber \\
    dz &=& \left( -2 \omega x - \gamma_e (1 + z) + \frac{\gamma_g}{2} (1 - z^2) \right) dt  \nonumber \\
       && \quad + \sqrt{\gamma_e} (-x (1 + z)) dW,
    \label{eq:Ito_form_y}
\end{eqnarray}
for the $y$-axis drive.
By taking ensemble average of Eq.\ref{eq:stochastic_master_post_select_normalized_appendix} or Eq.\ref{eq:Ito_form_x} and Eq.\ref{eq:Ito_form_y}, and assuming $\langle \rho(t)\, dW(t) \rangle = 0$, we obtain the normalized population similar to the one obtained from Eq.\ref{eq:Liouvilian diff}, i.e., the Liouvillain dynamics.

\section{Optimal Path Analysis.}
\label{appendix D}
Given an initial and final state of the non-Hermitian qubit, one can find the optimal path of the trajectories by extremizing the action $S$ associated with the trajectories. The action for each trajectory is defined as  
\begin{equation}
S = \int_0^T dt \, \Big( -p \cdot \dot{q} + \mathcal{H}(q,p,r) \Big),
\label{eq:action_definition}
\end{equation}
where  
\begin{equation}
\mathcal{H}(q,p,r) = p \cdot F[q,r] + G[q,r].
\label{eq:H_definition}
\end{equation}

Here, $q = [x,y,z]$ are the Bloch components, $p$ corresponds to the conjugate momentum, and $r$ is the measurement record. The function $F[q,r] = \dot{q}$ is given by Eqs.~\eqref{eq:trajectory_appebdix_sigmax_drive} or \eqref{eq:trajectory_appebdix_sigmay_drive}, and $G[q,r] = \mathrm{Tr}(\Sigma \rho + \rho \Sigma^\dagger)$.  

By extremizing the action, i.e., setting $\delta S = 0$, we obtain a set of constraint equations:  
\begin{equation}
\begin{cases}
\dot{q} = \dfrac{\partial \mathcal{H}}{\partial p}, \\[6pt]
\dot{p} = - \dfrac{\partial \mathcal{H}}{\partial q}, \\[6pt]
0 = \dfrac{\partial \mathcal{H}}{\partial r}.
\end{cases}
\label{eq:constraints}
\end{equation}

For the $\sigma_y$ drive and $\theta = 0$ measurement, we obtain the following equations for the Bloch vector:
\begin{equation}
\begin{aligned}
\dot{x} &= 2\omega z + \tfrac{\gamma}{2} x z + r \sqrt{\gamma_e}(1+z-x^2), \\
\dot{y} &= \tfrac{\gamma}{2} y z + \sqrt{\gamma_e}(-x y) r, \\
\dot{z} &= -2\omega x + \tfrac{\gamma}{2}(z^2 - 1) + \sqrt{\gamma_e}(z+1)(-x)r.
\label{eq:xdot}
\end{aligned}
\end{equation}

The conjugate momenta evolve as
\begin{equation}
\begin{aligned}
\dot{p}_x &= -\Big( p_x\!\left(\tfrac{\gamma}{2}z - 2r\sqrt{\gamma_e}x\right) 
+ p_y(-\sqrt{\gamma_e}ry)  \\ 
&\quad\;\; + p_z(-2\omega - \sqrt{\gamma_e}r(z+1)) 
+ r\sqrt{\gamma_e} \Big), \\
\dot{p}_y &= -\Big( p_y\!\left(\tfrac{\gamma}{2}z - r\sqrt{\gamma_e}x\right) \Big), 
\\
\dot{p}_z &= -\Big( p_x\!\left(2\omega + \tfrac{\gamma}{2}x + r\sqrt{\gamma_e}\right) 
+ p_y(\sqrt{\gamma_e}y)  \\ 
&\quad\;\; + p_z(\gamma z - r\sqrt{\gamma_e}) - \tfrac{\gamma}{2} \Big).
\label{eq:pzdot}
\end{aligned}
\end{equation}
The optimal measurement record $r$ is obtained self-consistently as  
\begin{align}
r &= p_x \sqrt{\gamma_e}(1+z-x^2) 
+ p_y \sqrt{\gamma_e}(-xy) \nonumber \\ 
&\quad\;\; + p_z \sqrt{\gamma_e}(z+1)(-x) 
+ \sqrt{\gamma_e} x .
\label{eq:r_optimal}
\end{align}
Thus, one can numerically solve for $x$, $y$, and $z$ for arbitrary initial and final states by substituting the optimized $r$ from Eq.~\eqref{eq:r_optimal} into Eqs.~\eqref{eq:xdot}--\eqref{eq:pzdot} and obtain the corresponding optimal path.

We can perform a one-dimensional phase-space analysis of the qubit optimal path by substituting 
$y = 0$, $x = \sin\theta$, and $z = \cos\theta$. Each trajectory corresponds to a pure state, with $\theta = 0 \rightarrow \ket{e}$ and $\theta = \pi \rightarrow \ket{f}$. 
The optimal path equations then reduce to
\begin{align}
r &= p \sqrt{\gamma_e}\,(1+\cos\theta) + \sqrt{\gamma_e}\,\sin\theta, \label{eq:r_1d} \\
\dot{\theta} &= 2\omega + \tfrac{\gamma}{2}\sin\theta 
+ r\sqrt{\gamma_e}\,(1+\cos\theta), \label{eq:theta_1d} \\
\dot{p} &= -\Big( 
p\big(\tfrac{\gamma}{2}\cos\theta - r\sqrt{\gamma_e}\sin\theta\big) 
+ r\sqrt{\gamma_e}\cos\theta + \tfrac{\gamma}{2}\sin\theta 
\Big). \label{eq:p_1d}
\end{align}

The corresponding Hamiltonian can be written as
\begin{equation}
\mathcal{H} = A(\theta)\,p^2 + B(\theta)\,p + C(\theta),
\label{eq:H_1d}
\end{equation}
with
\begin{equation}
\begin{aligned}
A(\theta) &= \tfrac{\gamma_e}{2}\,(1+\cos\theta)^2, \\
B(\theta) &= 2\omega + \Big(\tfrac{3}{2}\gamma_e - \tfrac{\gamma_g}{2}\Big)\sin\theta 
+ \tfrac{\gamma_e}{2}\sin(2\theta), \\
C(\theta) &= \tfrac{\gamma_e}{2}(1-\cos\theta) 
+ \tfrac{\gamma_e}{4}(1-\cos(2\theta)) \\
&\quad - \gamma_e - \tfrac{\gamma_g}{2}(1-\cos\theta).
\end{aligned}
\label{eq:ABC}
\end{equation}
This Hamiltonian $\mathcal{H}$ is a constant of motion $E$ for the optimal path.  
Consequently, one can plot the conjugate momentum $p(\theta, E)$ to visualize the qubit optimal trajectory in phase space. 


\bibliography{ref}

@article{BliasRevModPhys.93.025005,
  title = {Circuit quantum electrodynamics},
  author = {Blais, Alexandre and Grimsmo, Arne L. and Girvin, S. M. and Wallraff, Andreas},
  journal = {Rev. Mod. Phys.},
  volume = {93},
  issue = {2},
  pages = {025005},
  numpages = {72},
  year = {2021},
  publisher = {American Physical Society},
  doi = {10.1103/RevModPhys.93.025005},
  url = {https://link.aps.org/doi/10.1103/RevModPhys.93.025005}
}

@article{KochPhysRevA.76.042319,
  title = {Charge-insensitive qubit design derived from the Cooper pair box},
  author = {Koch, Jens and Yu, Terri M. and Gambetta, Jay and Houck, A. A. and Schuster, D. I. and Majer, J. and Blais, Alexandre and Devoret, M. H. and Girvin, S. M. and Schoelkopf, R. J.},
  journal = {Phys. Rev. A},
  volume = {76},
  issue = {4},
  pages = {042319},
  numpages = {19},
  year = {2007},
  doi = {10.1103/PhysRevA.76.042319},
  url = {https://link.aps.org/doi/10.1103/PhysRevA.76.042319}
}

@article{Cbender_1999,
  title = {PT-symmetric quantum mechanics },
  author = {Bender, Carl M and Boettcher, Stefan and Meisinger Peter N},
  journal = {J. Math. Phys.},
  volume = {40},
  issue = {},
  pages = { 2201–2229},
  numpages = {},
  year = {1999},
  doi = {https://doi.org/10.1063/1.532860},
  url = {https://pubs.aip.org/aip/jmp/article-abstract/40/5/2201/395777/symmetric-quantum-mechanics}
}

@article{Bender_2007,
doi = {10.1088/0034-4885/70/6/R03},
url = {https://doi.org/10.1088/0034-4885/70/6/R03},
year = {2007},
month = {may},
publisher = {},
volume = {70},
number = {6},
pages = {947},
author = {Bender, Carl M},
title = {Making sense of non-Hermitian Hamiltonians},
journal = {Reports on Progress in Physics},}

@article{PhysRevA_91_052113,
  title = {PT-symmetric Hamiltonians and their application in quantum information},
  author = {Croke, Sarah},
  journal = {Phys. Rev. A},
  volume = {91},
  issue = {5},
  pages = {052113},
  numpages = {12},
  year = {2015},
  month = {May},
  publisher = {American Physical Society},
  doi = {10.1103/PhysRevA.91.052113},
  url = {https://link.aps.org/doi/10.1103/PhysRevA.91.052113}
}

@article{heiss2012,
  author  = {W.~D.~Heiss},
  title   = {The physics of exceptional points},
  journal = {Journal of Physics A: Mathematical and Theoretical},
  year    = {2012},
  volume  = {45},
  number  = {44},
  pages   = {444016},
  doi     = {10.1088/1751-8113/45/44/444016},
  url     = {https://doi.org/10.1088/1751-8113/45/44/444016}
}

@article{Miri_2019,
doi = {10.1126/science.aar7709},
url = {https://www.science.org/doi/10.1126/science.aar7709},
year = {2019},
month = {},
publisher = {},
volume = {363},
number = {},
pages = {eaar7709},
author = { Miri, M A and Alu, Andrea  },
title = {Exceptional points in optics and photonics},
journal = {Science},}

@article{Ozdemir_2019,
doi = {https://doi.org/10.1038/s41563-019-0304-9},
url = {https://www.nature.com/articles/s41563-019-0304-9},
year = {2019},
month = {August},
publisher = {},
volume = {18},
number = {},
pages = {783–798},
author = { Ozdemir, S.K. and Rotter , S and Nori, F and Yang, L },
title = {Parity–time symmetry and exceptional points in photonics},
journal = {Nature Material},}

@Article{Peng2014,
author={Peng, Bo
and Ozdemir, cSahin Kaya
and Lei, Fuchuan
and Monifi, Faraz
and Gianfreda, Mariagiovanna
and Long, Gui Lu
and Fan, Shanhui
and Nori, Franco
and Bender, Carl M.
and Yang, Lan},
title={Parity--time-symmetric whispering-gallery microcavities},
journal={Nature Physics},
year={2014},
month={May},
day={01},
volume={10},
number={5},
pages={394-398},
issn={1745-2481},
doi={10.1038/nphys2927},
url={https://doi.org/10.1038/nphys2927}
}

@article{feng2017,
  title = {Non-Hermitian photonics based on parity–time symmetry},
  author = {Liang Feng and Ramy El-Ganainy and Li Ge},
  journal = {Nature Photonics},
  year = {2017},
  volume = {11},
  number = {12},
  pages = {752--762},
  isbn = {1749-4893},
  doi = {10.1038/s41566-017-0031-1},
  url = {https://doi.org/10.1038/s41566-017-0031-1}
}

@article{hodaei2014,
  title = {Parity-time–symmetric microring lasers},
  author = {Hossein Hodaei and Mohammad-Ali Miri and Matthias Heinrich and Demetrios N. Christodoulides and Mercedeh Khajavikhan},
  journal = {Science},
  booktitle = {Science},
  publisher = {American Association for the Advancement of Science},
  year = {2014},
  volume = {346},
  number = {6212},
  pages = {975--978},
  doi = {10.1126/science.1258480},
  url = {https://doi.org/10.1126/science.1258480}
}

@article{PhysRevLett.115.040402,
  title = {Observation of a Topological Transition in the Bulk of a Non-Hermitian System},
  author = {Zeuner, Julia M. and Rechtsman, Mikael C. and Plotnik, Yonatan and Lumer, Yaakov and Nolte, Stefan and Rudner, Mark S. and Segev, Mordechai and Szameit, Alexander},
  journal = {Phys. Rev. Lett.},
  volume = {115},
  issue = {4},
  pages = {040402},
  numpages = {5},
  year = {2015},
  month = {Jul},
  publisher = {American Physical Society},
  doi = {10.1103/PhysRevLett.115.040402},
  url = {https://link.aps.org/doi/10.1103/PhysRevLett.115.040402}
}

@article{xiao2017,
  title = {Observation of topological edge states in parity–time-symmetric quantum walks},
  author = {L. Xiao and X. Zhan and Z. H. Bian and K. K. Wang and X. Zhang and X. P. Wang and J. Li and K. Mochizuki and D. Kim and N. Kawakami and W. Yi and H. Obuse and B. C. Sanders and P. Xue},
  journal = {Nature Physics},
  year = {2017},
  volume = {13},
  number = {11},
  pages = {1117--1123},
  isbn = {1745-2481},
  doi = {10.1038/nphys4204},
  url = {https://www.nature.com/articles/nphys4204}}

@article{Peng,
  title = {Loss-induced suppression and revival of lasing},
  author = {B. Peng and S. K. Ozdemir and S. Rotter and H. Yilmaz and M. Liertzer and F. Monifi and C. M. Bender and F. Nori and L. Yang},
  journal = {Science},
  publisher = {American Association for the Advancement of Science},
  year = {2014},
  volume = {346},
  number = {6207},
  pages = {328--332},
  doi = {10.1126/science.1258004},
  url = {https://doi.org/10.1126/science.1258004}
}

@article{hodaei2017,
  title = {Enhanced sensitivity at higher-order exceptional points},
  author = {Hossein Hodaei and Absar U. Hassan and Steffen Wittek and Hipolito Garcia-Gracia and Ramy El-Ganainy and Demetrios N. Christodoulides and Mercedeh Khajavikhan},
  journal = {Nature},
  year = {2017},
  volume = {548},
  number = {7666},
  pages = {187--191},
  isbn = {1476-4687},
  doi = {10.1038/nature23280},
  url = {https://doi.org/10.1038/nature23280}
}

@article{xu2016,
  title = {Topological energy transfer in an optomechanical system with exceptional points},
  author = {H. Xu and D. Mason and Luyao Jiang and J. G. E. Harris},
  journal = {Nature},
  year = {2016},
  volume = {537},
  number = {7618},
  pages = {80--83},
  isbn = {1476-4687},
  doi = {10.1038/nature18604},
  url = {https://doi.org/10.1038/nature18604}
}

@article{zhang2017,
  title = {Observation of the exceptional point in cavity magnon-polaritons},
  author = {Dengke Zhang and Xiao-Qing Luo and Yi-Pu Wang and Tie-Fu Li and J. Q. You},
  journal = {Nature Communications},
  year = {2017},
  volume = {8},
  number = {1},
  pages = {1368},
  isbn = {2041-1723},
  doi = {10.1038/s41467-017-01634-w},
  url = {https://doi.org/10.1038/s41467-017-01634-w}
}

@article{PhysRevB.100.134505,
  title = {Exceptional points in tunable superconducting resonators},
  author = {Partanen, Matti and Goetz, Jan and Tan, Kuan Yen and Kohvakka, Kassius and Sevriuk, Vasilii and Lake, Russell E. and Kokkoniemi, Roope and Ikonen, Joni and Hazra, Dibyendu and Makinen, Akseli and Hyyppa, Eric and Gronberg, Leif and Vesterinen, Visa and Silveri, Matti and Mottonen, Mikko},
  journal = {Phys. Rev. B},
  volume = {100},
  issue = {13},
  pages = {134505},
  numpages = {17},
  year = {2019},
  month = {Oct},
  publisher = {American Physical Society},
  doi = {10.1103/PhysRevB.100.134505},
  url = {https://link.aps.org/doi/10.1103/PhysRevB.100.134505}
}

@article{shi2016,
  title = {Accessing the exceptional points of parity-time symmetric acoustics},
  author = {Chengzhi Shi and Marc Dubois and Yun Chen and Lei Cheng and Hamidreza Ramezani and Yuan Wang and Xiang Zhang},
  journal = {Nature Communications},
  year = {2016},
  volume = {7},
  number = {1},
  pages = {11110},
  isbn = {2041-1723},
  doi = {10.1038/ncomms11110},
  url = {https://doi.org/10.1038/ncomms11110}
}

@article{chen2017,
  title = {Exceptional points enhance sensing in an optical microcavity},
  author = {Weijian Chen and Sahin Kaya Ozdemir and Guangming Zhao and Jan Wiersig and Lan Yang},
  journal = {Nature},
  year = {2017},
  volume = {548},
  number = {7666},
  pages = {192--196},
  isbn = {1476-4687},
  doi = {10.1038/nature23281},
  url = {https://doi.org/10.1038/nature23281}
}

@article{lau2018,
  title = {Fundamental limits and non-reciprocal approaches in non-Hermitian quantum sensing},
  author = {Hoi-Kwan Lau and Aashish A. Clerk},
  journal = {Nature Communications},
  year = {2018},
  volume = {9},
  number = {1},
  pages = {4320},
  isbn = {2041-1723},
  doi = {10.1038/s41467-018-06477-7},
  url = {https://doi.org/10.1038/s41467-018-06477-7}
}

@article{naghiloo2019,
  title = {Quantum state tomography across the exceptional point in a single dissipative qubit},
  author = {M. Naghiloo and M. Abbasi and Yogesh N. Joglekar and K. W. Murch},
  journal = {Nature Physics},
  year = {2019},
  volume = {15},
  number = {12},
  pages = {1232--1236},
  isbn = {1745-2481},
  doi = {10.1038/s41567-019-0652-z},
  url = {https://doi.org/10.1038/s41567-019-0652-z}
}

@article{Chen2021,
  title = {Quantum Jumps in the Non-Hermitian Dynamics of a Superconducting Qubit},
  author = {Chen, Weijian and Abbasi, Maryam and Joglekar, Yogesh N. and Murch, Kater W.},
  journal = {Phys. Rev. Lett.},
  volume = {127},
  issue = {14},
  pages = {140504},
  numpages = {6},
  year = {2021},
  month = {Sep},
  publisher = {American Physical Society},
  doi = {10.1103/PhysRevLett.127.140504},
  url = {https://link.aps.org/doi/10.1103/PhysRevLett.127.140504}
}

@article{Ming2019,
  title = {Quantum exceptional points of non-Hermitian Hamiltonians and Liouvillians: The effects of quantum jumps},
  author = {Minganti, Fabrizio and Miranowicz, Adam and Chhajlany, Ravindra W. and Nori, Franco},
  journal = {Phys. Rev. A},
  volume = {100},
  issue = {6},
  pages = {062131},
  numpages = {17},
  year = {2019},
  month = {Dec},
  publisher = {American Physical Society},
  doi = {10.1103/PhysRevA.100.062131},
  url = {https://link.aps.org/doi/10.1103/PhysRevA.100.062131}
}

@article{PhysRevA.101.062112,
  title = {Hybrid-Liouvillian formalism connecting exceptional points of non-Hermitian Hamiltonians and Liouvillians via postselection of quantum trajectories},
  author = {Minganti, Fabrizio and Miranowicz, Adam and Chhajlany, Ravindra W. and Arkhipov, Ievgen I. and Nori, Franco},
  journal = {Phys. Rev. A},
  volume = {101},
  issue = {6},
  pages = {062112},
  numpages = {14},
  year = {2020},
  month = {Jun},
  publisher = {American Physical Society},
  doi = {10.1103/PhysRevA.101.062112},
  url = {https://link.aps.org/doi/10.1103/PhysRevA.101.062112}
}

@article{lewalle2020,
  title = {Measuring fluorescence to track a quantum emitter's state: a theory review},
  author = {Philippe Lewalle and Sreenath K. Manikandan and Cyril Elouard and Andrew N. Jordan},
  journal = {Contemporary Physics},
  booktitle = {Contemporary Physics},
  publisher = {Taylor \& Francis},
  year = {2020},
  volume = {61},
  number = {1},
  pages = {26--50},
  isbn = {0010-7514},
  doi = {10.1080/00107514.2020.1747201},
  url = {https://www.tandfonline.com/doi/citedby/10.1080/00107514.2020.1747201?scroll=top&needAccess=true}
}

@article{lewallethesis,
  author  = {P.~Lewalle},
  title   = {Quantum Trajectories and Their Extremal-Probability Paths: New Phenomena and Applications},
  journal = {Thesis}
}

@article{jordan2016,
  title = {Anatomy of fluorescence: quantum trajectory statistics from continuously measuring spontaneous emission},
  author = {Andrew N. Jordan and Areeya Chantasri and Pierre Rouchon and Benjamin Huard},
  journal = {Quantum Studies: Mathematics and Foundations},
  year = {2016},
  volume = {3},
  number = {3},
  pages = {237--263},
  isbn = {2196-5617},
  doi = {10.1007/s40509-016-0075-9},
  url = {https://doi.org/10.1007/s40509-016-0075-9}
}

@article{PhysRevA.96.053807,
  title = {Quantum caustics in resonance-fluorescence trajectories},
  author = {Naghiloo, M. and Tan, D. and Harrington, P. M. and Lewalle, P. and Jordan, A. N. and Murch, K. W.},
  journal = {Phys. Rev. A},
  volume = {96},
  issue = {5},
  pages = {053807},
  numpages = {11},
  year = {2017},
  month = {Nov},
  publisher = {American Physical Society},
  doi = {10.1103/PhysRevA.96.053807},
  url = {https://link.aps.org/doi/10.1103/PhysRevA.96.053807}
}

@article{PhysRevX.6.011002,
  title = {Observing Quantum State Diffusion by Heterodyne Detection of Fluorescence},
  author = {Campagne-Ibarcq, P. and Six, P. and Bretheau, L. and Sarlette, A. and Mirrahimi, M. and Rouchon, P. and Huard, B.},
  journal = {Phys. Rev. X},
  volume = {6},
  issue = {1},
  pages = {011002},
  numpages = {7},
  year = {2016},
  month = {Jan},
  publisher = {American Physical Society},
  doi = {10.1103/PhysRevX.6.011002},
  url = {https://link.aps.org/doi/10.1103/PhysRevX.6.011002}
}

@article{naghiloo2016,
  title = {Mapping quantum state dynamics in spontaneous emission},
  author = {M. Naghiloo and N. Foroozani and D. Tan and A. Jadbabaie and K. W. Murch},
  journal = {Nature Communications},
  year = {2016},
  volume = {7},
  number = {1},
  pages = {11527},
  isbn = {2041-1723},
  doi = {10.1038/ncomms11527},
  url = {https://doi.org/10.1038/ncomms11527}
}

@article{PhysRevA.96.022104,
  title = {Homodyne monitoring of postselected decay},
  author = {Tan, D. and Foroozani, N. and Naghiloo, M. and Kiilerich, A. H. and Molmer, K. and Murch, K. W.},
  journal = {Phys. Rev. A},
  volume = {96},
  issue = {2},
  pages = {022104},
  numpages = {9},
  year = {2017},
  month = {Aug},
  publisher = {American Physical Society},
  doi = {10.1103/PhysRevA.96.022104},
  url = {https://link.aps.org/doi/10.1103/PhysRevA.96.022104}
}

@article{ficheux2018,
  title = {Dynamics of a qubit while simultaneously monitoring its relaxation and dephasing},
  author = {Q. Ficheux and S. Jezouin and Z. Leghtas and B. Huard},
  journal = {Nature Communications},
  year = {2018},
  volume = {9},
  number = {1},
  pages = {1926},
  isbn = {2041-1723},
  doi = {10.1038/s41467-018-04372-9},
  url = {https://doi.org/10.1038/s41467-018-04372-9}
}

@article{hmwiseman1996,
  author  = {H.~M.~Wiseman},
  title   = {Quantum trajectories and quantum measurement theory},
  journal = {Quantum and Semiclassical Optics: Journal of the European Optical Society Part B},
  year    = {1996},
  volume  = {8},
  number  = {1},
  pages   = {205},
  doi     = {10.1088/1355-5111/8/1/015},
  url     = {https://doi.org/10.1088/1355-5111/8/1/015}
}

@article{PhysRevLett.70.548,
  title = {Quantum theory of optical feedback via homodyne detection},
  author = {Wiseman, H. M. and Milburn, G. J.},
  journal = {Phys. Rev. Lett.},
  volume = {70},
  issue = {5},
  pages = {548--551},
  numpages = {0},
  year = {1993},
  month = {Feb},
  publisher = {American Physical Society},
  doi = {10.1103/PhysRevLett.70.548},
  url = {https://link.aps.org/doi/10.1103/PhysRevLett.70.548}
}

@article{weber2014,
  title = {Mapping the optimal route between two quantum states},
  author = {S. J. Weber and A. Chantasri and J. Dressel and A. N. Jordan and K. W. Murch and I. Siddiqi},
  journal = {Nature},
  year = {2014},
  volume = {511},
  number = {7511},
  pages = {570--573},
  isbn = {1476-4687},
  doi = {10.1038/nature13559},
  url = {https://doi.org/10.1038/nature13559}
}

@article{jordan2013,
  title = {Watching the wavefunction collapse},
  author = {Andrew N. Jordan},
  journal = {Nature},
  year = {2013},
  volume = {502},
  number = {7470},
  pages = {177--178},
  isbn = {1476-4687},
  doi = {10.1038/502177a},
  url = {https://doi.org/10.1038/502177a}
}

@article{mlmer1993,
  title = {Monte Carlo wave-function method in quantum optics},
  author = {Klaus Mølmer and Yvan Castin and Jean Dalibard},
  journal = {J. Opt. Soc. Am. B},
  booktitle = {Journal of the Optical Society of America B},
  publisher = {Optica Publishing Group},
  year = {1993},
  volume = {10},
  number = {3},
  pages = {524--538},
  doi = {10.1364/JOSAB.10.000524},
  url = {https://opg.optica.org/josab/abstract.cfm?URI=josab-10-3-524}
}

@article{PhysRevA.89.023827,
  title = {Stochastic excitation during the decay of a two-level emitter subject to homodyne and heterodyne detection},
  author = {Bolund, Anders and M\o{}lmer, Klaus},
  journal = {Phys. Rev. A},
  volume = {89},
  issue = {2},
  pages = {023827},
  numpages = {9},
  year = {2014},
  month = {Feb},
  publisher = {American Physical Society},
  doi = {10.1103/PhysRevA.89.023827},
  url = {https://link.aps.org/doi/10.1103/PhysRevA.89.023827}
}

@article{jacobs2006,
  title = {A straightforward introduction to continuous quantum measurement},
  author = {Kurt Jacobs and Daniel A. Steck},
  journal = {Contemporary Physics},
  booktitle = {Contemporary Physics},
  publisher = {Taylor \& Francis},
  year = {2006},
  volume = {47},
  number = {5},
  pages = {279--303},
  isbn = {0010-7514},
  doi = {10.1080/00107510601101934},
  url = {https://doi.org/10.1080/00107510601101934}
}

@article{PhysRevA.94.042326,
  title = {Quantum Bayesian approach to circuit QED measurement with moderate bandwidth},
  author = {Korotkov, Alexander N.},
  journal = {Phys. Rev. A},
  volume = {94},
  issue = {4},
  pages = {042326},
  numpages = {24},
  year = {2016},
  month = {Oct},
  publisher = {American Physical Society},
  doi = {10.1103/PhysRevA.94.042326},
  url = {https://link.aps.org/doi/10.1103/PhysRevA.94.042326}
}

@article{murch2013,
  title = {Observing single quantum trajectories of a superconducting quantum bit},
  author = {K. W. Murch and S. J. Weber and C. Macklin and I. Siddiqi},
  journal = {Nature},
  year = {2013},
  volume = {502},
  number = {7470},
  pages = {211--214},
  isbn = {1476-4687},
  doi = {10.1038/nature12539},
  url = {https://doi.org/10.1038/nature12539}
}

@article{minev2019,
  title = {To catch and reverse a quantum jump mid-flight},
  author = {Z. K. Minev and S. O. Mundhada and S. Shankar and P. Reinhold and R. Gutiérrez-Jáuregui and R. J. Schoelkopf and M. Mirrahimi and H. J. Carmichael and M. H. Devoret},
  journal = {Nature},
  year = {2019},
  volume = {570},
  number = {7760},
  pages = {200--204},
  isbn = {1476-4687},
  doi = {10.1038/s41586-019-1287-z},
  url = {https://doi.org/10.1038/s41586-019-1287-z}
}

@article{PhysRevA.77.012112,
  title = {Quantum trajectory approach to circuit QED: Quantum jumps and the Zeno effect},
  author = {Gambetta, Jay and Blais, Alexandre and Boissonneault, M. and Houck, A. A. and Schuster, D. I. and Girvin, S. M.},
  journal = {Phys. Rev. A},
  volume = {77},
  issue = {1},
  pages = {012112},
  numpages = {18},
  year = {2008},
  month = {Jan},
  publisher = {American Physical Society},
  doi = {10.1103/PhysRevA.77.012112},
  url = {https://link.aps.org/doi/10.1103/PhysRevA.77.012112}
}

@article{PhysRevLett.52.1657,
  title = {Quantum Measurements and Stochastic Processes},
  author = {Gisin, N.},
  journal = {Phys. Rev. Lett.},
  volume = {52},
  issue = {19},
  pages = {1657--1660},
  numpages = {0},
  year = {1984},
  month = {May},
  publisher = {American Physical Society},
  doi = {10.1103/PhysRevLett.52.1657},
  url = {https://link.aps.org/doi/10.1103/PhysRevLett.52.1657}
}

@article{PhysRevA.34.1642,
  title = {Measurement theory and stochastic differential equations in quantum mechanics},
  author = {Barchielli, Alberto},
  journal = {Phys. Rev. A},
  volume = {34},
  issue = {3},
  pages = {1642--1649},
  numpages = {0},
  year = {1986},
  month = {Sep},
  publisher = {American Physical Society},
  doi = {10.1103/PhysRevA.34.1642},
  url = {https://link.aps.org/doi/10.1103/PhysRevA.34.1642}
}

@book{wiseman2009,
  title = {Quantum Measurement and Control},
  author = {Howard M. Wiseman and Gerard J. Milburn},
  publisher = {Cambridge University Press},
  address = {Cambridge},
  year = {2009},
  isbn = {9780521804424},
  doi = {DOI: 10.1017/CBO9780511813948},
  url = {https://www.cambridge.org/core/product/F78F445CD9AF00B10593405E9BAC6B9F}
}

@book{steck2007quantum,
  title={Quantum and Atom Optics},
  author={Steck, D.A.},
  url={https://books.google.co.in/books?id=bc9TMwEACAAJ},
  year={2007}
}

@article{PhysRevA.88.042110,
  title = {Action principle for continuous quantum measurement},
  author = {Chantasri, A. and Dressel, J. and Jordan, A. N.},
  journal = {Phys. Rev. A},
  volume = {88},
  issue = {4},
  pages = {042110},
  numpages = {7},
  year = {2013},
  month = {Oct},
  publisher = {American Physical Society},
  doi = {10.1103/PhysRevA.88.042110},
  url = {https://link.aps.org/doi/10.1103/PhysRevA.88.042110}
}

@article{PhysRevA.92.032125,
  title = {Stochastic path-integral formalism for continuous quantum measurement},
  author = {Chantasri, Areeya and Jordan, Andrew N.},
  journal = {Phys. Rev. A},
  volume = {92},
  issue = {3},
  pages = {032125},
  numpages = {22},
  year = {2015},
  month = {Sep},
  publisher = {American Physical Society},
  doi = {10.1103/PhysRevA.92.032125},
  url = {https://link.aps.org/doi/10.1103/PhysRevA.92.032125}
}

@article{PhysRevA.95.042126,
  title = {Prediction and characterization of multiple extremal paths in continuously monitored qubits},
  author = {Lewalle, Philippe and Chantasri, Areeya and Jordan, Andrew N.},
  journal = {Phys. Rev. A},
  volume = {95},
  issue = {4},
  pages = {042126},
  numpages = {18},
  year = {2017},
  month = {Apr},
  publisher = {American Physical Society},
  doi = {10.1103/PhysRevA.95.042126},
  url = {https://link.aps.org/doi/10.1103/PhysRevA.95.042126}
}

@article{PRXQuantum.3.010327,
  title = {Stochastic Path-Integral Analysis of the Continuously Monitored Quantum Harmonic Oscillator},
  author = {Karmakar, Tathagata and Lewalle, Philippe and Jordan, Andrew N.},
  journal = {PRX Quantum},
  volume = {3},
  issue = {1},
  pages = {010327},
  numpages = {18},
  year = {2022},
  month = {Feb},
  publisher = {American Physical Society},
  doi = {10.1103/PRXQuantum.3.010327},
  url = {https://link.aps.org/doi/10.1103/PRXQuantum.3.010327}
}

@article{PhysRevA.98.012141,
  title = {Chaos in continuously monitored quantum systems: An optimal-path approach},
  author = {Lewalle, Philippe and Steinmetz, John and Jordan, Andrew N.},
  journal = {Phys. Rev. A},
  volume = {98},
  issue = {1},
  pages = {012141},
  numpages = {21},
  year = {2018},
  month = {Jul},
  publisher = {American Physical Society},
  doi = {10.1103/PhysRevA.98.012141},
  url = {https://link.aps.org/doi/10.1103/PhysRevA.98.012141}
}

@article{PhysRevResearch.6.L032057,
  title = {Telling different unravelings apart via nonlinear quantum-trajectory averages},
  author = {Pinol, Eloy and Mavrogordatos, Th. K. and Keys, Dustin and Veyron, Romain and Sierant, Piotr and Angel Garcia-March, Miguel and Grandi, Samuele and Mitchell, Morgan W. and Wehr, Jan and Lewenstein, Maciej},
  journal = {Phys. Rev. Res.},
  volume = {6},
  issue = {3},
  pages = {L032057},
  numpages = {8},
  year = {2024},
  month = {Sep},
  publisher = {American Physical Society},
  doi = {10.1103/PhysRevResearch.6.L032057},
  url = {https://link.aps.org/doi/10.1103/PhysRevResearch.6.L032057}
}

@misc{etde_21197676,
title = {Quantum trajectories and measurements in continuous time. The diffusive case},
author = {Barchielli, Alberto and Gregoratti, Matteo},
doi = {10.1007/978-3-642-01298-3},
place = {Germany},
year = {2009},
month = {Jul}
}

@article{PhysRevLett.123.163601,
  title = {Observing and Verifying the Quantum Trajectory of a Mechanical Resonator},
  author = {Rossi, Massimiliano and Mason, David and Chen, Junxin and Schliesser, Albert},
  journal = {Phys. Rev. Lett.},
  volume = {123},
  issue = {16},
  pages = {163601},
  numpages = {6},
  year = {2019},
  month = {Oct},
  publisher = {American Physical Society},
  doi = {10.1103/PhysRevLett.123.163601},
  url = {https://link.aps.org/doi/10.1103/PhysRevLett.123.163601}
}

@misc{Breuer_lecture_notes,
  author       = {H.-P. Breuer},
  title        = {Lecture Notes on Open Quantum Systems},
  institution  = {University of Freiburg},
  note         = {Lecture notes}
}

@book{RivasHuelga2012,
  author    = {Ángel Rivas and Susana F. Huelga},
  title     = {Open Quantum Systems: An Introduction},
  publisher = {Springer},
  year      = {2012},
  series    = {SpringerBriefs in Physics},
  doi       = {10.1007/978-3-642-23354-8}
}

@book{Carmichael1993,
  author    = {H. J. Carmichael},
  title     = {An Open Systems Approach to Quantum Optics},
  publisher = {Springer},
  year      = {1993},
  series    = {Lecture Notes in Physics Monographs},
  volume    = {18},
  doi       = {10.1007/978-3-540-47620-7}
}

@article{Lami_2024,
doi = {10.1088/1367-2630/ad1f0a},
url = {https://doi.org/10.1088/1367-2630/ad1f0a},
year = {2024},
month = {feb},
publisher = {IOP Publishing},
volume = {26},
number = {2},
pages = {023041},
author = {Lami, Guglielmo and Santini, Alessandro and Collura, Mario},
title = {Continuously monitored quantum systems beyond Lindblad dynamics},
journal = {New Journal of Physics}
}

\end{document}